\documentclass[12pt]{article}
\pdfoutput=1

\usepackage[utf8]{inputenc}
\usepackage{amsmath,amssymb,mathtools}
\usepackage{graphicx}
\usepackage[hidelinks, pageanchor=false]{hyperref}
\usepackage{xcolor}
\usepackage{multirow}
\usepackage{xspace}
\usepackage[
  a4paper,
  twoside=false,
  left=1.7cm,
  right=1.7cm,
  top=2.5cm,
  bottom=3cm
]{geometry}

\allowdisplaybreaks
\DeclareRobustCommand{\nn}{\nonumber} 
\DeclareRobustCommand{\alphas}{\ensuremath{\alpha_{\mathrm{s}}}\xspace} 
\DeclareRobustCommand{\as}{\alphas} 
\DeclareRobustCommand{\asOpi}{\ensuremath{\left(\frac{\as}{\pi}\right)}} 
\DeclareRobustCommand{\mur}{\ensuremath{\mu_{\mathrm{R}}}\xspace}
\DeclareRobustCommand{\muR}{\mur}
\DeclareRobustCommand{\muf}{\ensuremath{\mu_{\mathrm{F}}}\xspace}
\DeclareRobustCommand{\muF}{\muf}
\DeclareRobustCommand{\qt}{\ensuremath{q_T}\xspace} 
\DeclareRobustCommand{\qtcut}{\ensuremath{q_T^\mathrm{cut}}\xspace} 

\DeclareRobustCommand{\rd}{\ensuremath{\mathrm{d}}}
\DeclareRobustCommand{\re}{\ensuremath{\mathrm{e}}}
\DeclareRobustCommand{\ri}{\ensuremath{\mathrm{i}}\xspace}
\DeclareRobustCommand{\cO}{\ensuremath{\mathcal{O}}}
\DeclareRobustCommand{\cG}{\ensuremath{\mathcal{G}}}
\DeclareRobustCommand{\cH}{\ensuremath{\mathcal{H}}}
\DeclareRobustCommand{\obs}{\cO\xspace}

\DeclareRobustCommand{\jets}{\text{jet(s)}\xspace}
\DeclareRobustCommand{\CT}{\text{CT}\xspace}
\DeclareRobustCommand{\fin}{\text{(fin.)}\xspace}
\DeclareRobustCommand{\tot}{\text{(tot.)}\xspace}
\DeclareRobustCommand{\LO}{\text{LO}\xspace}
\DeclareRobustCommand{\NLO}{\text{NLO}\xspace}
\DeclareRobustCommand{\NNLO}{\text{NNLO}\xspace}
\DeclareRobustCommand{\N}[1]{\ensuremath{\text{N}^{#1}}} 

\begin{document} 
\begin{titlepage}
\renewcommand{\thefootnote}{\fnsymbol{footnote}}
\begin{flushright}
ZU-TH 27/18, 
IPPP/18/67, 
CERN-TH-2018-167
\end{flushright}
\par \vspace{10mm}
\begin{center}
{\Large \bf
Higgs boson production at the LHC\\
\vskip .3cm
using the $\qt$ subtraction formalism at N$^\text{3}$LO QCD}
\end{center}
\par \vspace{2mm}
\begin{center}
{\bf Leandro Cieri${}^{(a,b)}$, Xuan Chen${}^{(b)}$, Thomas Gehrmann${}^{(b)}$,\\  E.W.N. Glover${}^{(c)}$ and Alexander Huss${}^{(d)}$ }

\vspace{5mm}

$^{(a)}$ INFN, Sezione di Milano-Bicocca,
Piazza della Scienza 3, I-20126 Milano, Italy

$^{(b)}$ Physik-Institut, Universit\"at Z\"urich, 
CH-8057 Zurich, Switzerland

$^{(c)}$ Institute for Particle Physics Phenomenology, Durham University, Durham, DH1 3LE, UK

$^{(d)}$ Theoretical Physics Department, CERN, 1211 Geneva 23, Switzerland

\vspace{5mm}

\end{center}

\par \vspace{2mm}
\begin{center} {\large \bf Abstract} \end{center}
\begin{quote}
\pretolerance 10000

We consider higher-order QCD corrections to Higgs boson production through gluon--gluon fusion in the large top quark mass limit in hadron collisions. We extend the transverse-momentum ($\qt$) subtraction method to next-to-next-to-next-to-leading order (\N3\LO) and combine it with the NNLO Higgs-plus-jet calculation to numerically compute differential infrared-safe observables at \N3\LO for Higgs boson production in gluon fusion. To cancel the infrared divergences, we exploit the universal behaviour of the associated $\qt$ distributions in the small-$\qt$ region. We document all the necessary ingredients of the transverse-momentum subtraction method up to \N3\LO. The missing third-order collinear functions, which contribute only at $\qt=0$, are approximated using a prescription which uses the known result for the total Higgs boson cross section at this order. As a first application of the third-order $\qt$ subtraction method, we present the \N3\LO rapidity distribution of the Higgs boson at the LHC.

\end{quote}
\end{titlepage}
\setcounter{footnote}{1}
\renewcommand{\thefootnote}{\fnsymbol{footnote}}

\section{Introduction}
\label{sec:intro}

The most straightforward and successful (as well as systematically improvable) approach to calculations for processes at high-momentum scales $M$ in QCD is a perturbative expansion in the strong coupling $\as(M^{2})$. Cross sections are written as a series expansion in the parameter $\as$ and an improvement in accuracy is obtained by calculating an increasing number of coefficients in the series. Until a few years ago, the standard for such calculations was next-to-leading order (NLO) accuracy. Recent years have seen a number of next-to-next-to-leading order (NNLO) results for many important processes of interest, such that the emerging standard for precision calculations relevant for LHC phenomenology is the second non-trivial order in the strong coupling $\as$.

Reducing the theoretical uncertainties remains one of the main motivations for the extension from NLO to NNLO accuracy. This is particularly relevant in two distinct situations. Firstly, NNLO corrections are mandatory for those processes where NLO corrections are comparable in size to the leading order (LO) contribution, both to establish the convergence of the perturbative expansion and to obtain reliable predictions. Secondly, many benchmark processes demand theoretical predictions with the highest possible precision to be able to fully exploit the extraordinary experimental precision that is achievable for this class of processes. Such ``standard candles'' are not only indispensable tools in detector calibration but also allow for a precise extraction of Standard Model (SM) parameters and parton distribution functions (PDFs).

Extending the perturbative accuracy of QCD calculations to one order higher implies developing new methods and techniques to achieve the cancellation of infrared (IR) divergences that appear at intermediate steps of the calculations. The past few years have witnessed a great development in NNLO subtraction prescriptions. The transverse momentum ($\qt$) subtraction method~\cite{Catani:2007vq,Bozzi:2005wk,Bonciani:2015sha}, the $N$-jettiness subtraction~\cite{Boughezal:2015eha, Gaunt:2015pea}, projection-to-Born~\cite{Cacciari:2015jma}, residue subtraction~\cite{Czakon:2011ve,Boughezal:2011jf}, and the antenna subtraction method~\cite{Antenna:method} have all been successfully applied for LHC phenomenology. 

However, in view of the impressive and continuously improving quality of the measurements performed at the LHC, even NNLO accuracy is in some cases not sufficient to match the demands of the LHC data. Typically, these are processes in which the size of the NLO corrections are comparable with the LO, and where the NNLO corrections still exhibit large effects such that the size of the theoretical uncertainties remains larger than the experimental uncertainties. 

This motivated a new theoretical effort to go beyond NNLO to include the next perturbative order: the next-to-next-to-next-to-leading order (\N3\LO). Sum rules, branching fractions~\cite{Chetyrkin:1994js} and deep inelastic structure functions~\cite{Vermaseren:2005qc} have been known to this order for quite some time. At present, the only hadron collider observables for which \N3\LO QCD corrections have been calculated are the total cross section for Higgs boson  production in gluon fusion~\cite{Anastasiou:2015ema,Mistlberger:2018etf} and in vector boson 
fusion~\cite{Dreyer:2016oyx}. First steps have been taken towards more differential observables by computing several \N3\LO threshold expansion terms to the Higgs boson rapidity distribution in gluon fusion~\cite{Dulat:2017prg,Dulat:2018bfe}. Moreover, the projection-to-Born method has been most recently extended to compute fully differential distributions to \N3\LO, with a proof-of-principle calculation~\cite{Currie:2018fgr} of jet production in deep inelastic scattering. 

In this paper we extend the $\qt$ subtraction method at \N3\LO  to compute Higgs boson production differentially in the Higgs boson rapidity at \N3\LO accuracy. The paper is organized as follows: in Sec.~\ref{sec:forma} we recall briefly the main ideas of the $\qt$ subtraction formalism and we present the necessary ingredients up to  \N3\LO, specifying which elements are known analytically and identifying the missing coefficients at \N3\LO. In Sec.~\ref{sec:numforCN3} we present a prescription for approximating the missing collinear functions at \N3\LO based on the unitarity property of the integral of the transverse momentum distribution. In Sec.~\ref{sec:rapgen}, we apply the $\qt$ subtraction formalism at \N3\LO to produce differential distributions in the rapidity of the Higgs boson. To validate our approach, Sec.~\ref{sec:NNLOrap} quantifies the quality of the approximations by repeating them at NNLO, where all of the ingredients to $\qt$ subtraction are known. We assess the magnitude of different sources of systematic uncertainties at  \N3\LO in Sec.~\ref{sec:N3LOrap}, yielding final results for the \N3\LO Higgs boson rapidity distribution and the associated theoretical uncertainty in Sec.~\ref{sec:results}. Finally, in Sec.~\ref{Sec:conclu} we summarize our results.

\section{The \texorpdfstring{$\qt$}{qT} subtraction formalism at \texorpdfstring{N${}^\text{3}$LO}{N3LO}}
\label{sec:forma}

This section is devoted to present briefly the transverse-momentum subtraction formalism to \N3\LO in perturbative QCD. The method is illustrated in its general form and special attention is paid to the case of Higgs boson production through gluon--gluon fusion. The $\qt$ subtraction formalism presented in this section is the third-order extension of the subtraction method originally proposed in Refs.~\cite{Catani:2007vq,Bozzi:2005wk,Bonciani:2015sha}.

We consider the inclusive hard scattering reaction 
\begin{equation}
  h_1(p_1)+h_2(p_2)\to F(\{q_i\})+X\, ,
  \label{class}
\end{equation}
where $h_1$ and $h_2$ denote the two hadrons which collide with momenta $p_1$ and $p_2$ producing the identified colourless final-state system $F$, accompanied by an arbitrary and undetected final state $X$. The colliding hadrons have centre-of-mass energy $\sqrt s$, and are treated as massless particles 
$$s= (p_1+p_2)^2 = 2p_1\cdot p_2 \;.$$ 
The observed final state $F$ consists of a generic system of non-QCD partons composed of \emph{one or more} colour singlet particles (such as vector bosons, photons, Higgs bosons, Drell--Yan (DY) lepton pairs and so forth) with momenta $q_i^{\mu}$ $(i=3,4,5,\dots)$. The total momentum of the system $F$ is denoted by 
$$q^{\mu}=\sum_i q_i^\mu \;,$$
and the kinematics of the system can be expressed in terms of the total invariant mass $M$, 
$$M^2=q^2 \;,$$ 
the transverse momentum $\qt$ with respect to the direction of the colliding hadrons (omitting the azimuthal dependence), and the rapidity in the centre-of-mass system of the hadronic collision, $Y$, 
$$Y = \frac{1}{2} \ln \left(\frac{p_2\cdot q}{p_1\cdot q}\right) \;.$$ 
The fully differential hadronic cross section can therefore be written as 
\begin{align}
  \frac{\rd\sigma^{F}}{\rd\qt^2\rd M^2\rd Y}
  &=
  \int_0^1\rd \xi_1 \int_0^1\rd \xi_2 \; 
  \frac{\rd\hat{\sigma}^{F}_{ab}(\xi_1 p_1, \xi_2 p_2)}{\rd\qt^2\rd M^2\rd Y} \;
  f_{a/h_1}(\xi_1,\muf) \; f_{b/h_2}(\xi_2,\muf) \;,
  \label{eq:hadXS_diff}
\end{align}
where $\rd\hat{\sigma}_{ab}$ is the differential partonic cross section, $\xi_1$, $\xi_2$ are the partonic momentum fractions and $f_{c/h}$ the distribution function for finding parton $c$ in hadron $h$.
Since $F$ is colourless, the LO partonic cross section can be either initiated by 
$q\bar{q}$ annihilation, as in the case of the Drell--Yan process, or by gluon--gluon fusion, as in the case of Higgs boson production.
In the case of the Born cross section, the kinematics of the colour-neutral system $F$ is fully constrained such that
\begin{align}
  \frac{\rd\hat{\sigma}^{F}_{\LO;ab}(\xi_1 p_1, \xi_2 p_2)}{\rd\qt^2\rd M^2\rd Y}
  &=
  \delta_{c\,a} \, \delta_{\bar{c}\,b} \, 
  \delta(\qt^2) \,
  \delta\bigl(M^2-\xi_1 \xi_2 s\bigr) \,
  \delta\bigl(Y-\ln(\xi_1/\xi_2)/2\bigr) \,
  \rd\hat{\sigma}^{F}_{\LO;c\bar{c}}(\xi_1 p_1, \xi_2 p_2)
  \nn\\
  &=
  \delta_{c\,a} \, \delta_{\bar{c}\,b} \, 
  \delta(\qt^2) \,
  \delta\Bigl(\xi_1-\frac{M}{\sqrt{s}}\;\re^{+Y}\Bigr) \,
  \delta\Bigl(\xi_2-\frac{M}{\sqrt{s}}\;\re^{-Y}\Bigr) \,
  \frac{1}{s} \;
  \rd\hat{\sigma}^{F}_{\LO;c\bar{c}}(\xi_1 p_1, \xi_2 p_2) \;.
  \label{eq:LOpartXS_diff}
\end{align}

In order to explain the basic idea  of the subtraction formalism, we first notice that at LO, the transverse momentum $\qt$ of the final state system $F$ is identically zero. Therefore, as long as $\qt>0$, the \N{n}\LO QCD contributions (with $n\geq1$) are given by the \N{n-1}\LO QCD contributions to the $F+\jets$ final state. Consequently, if $\qt>0$ we have:
\begin{equation}
  \left. \frac{\rd\sigma^{F}_{\N{n}\LO}}{\rd\qt^2\rd M^2\rd Y} \right\rvert_{\qt>0}
  \equiv 
  \frac{\rd\sigma^{F+\jets}_{\N{n-1}\LO}}{\rd\qt^2\rd M^2\rd Y} \;,
  \label{Eq:first}
\end{equation}
where the notation \N{n}\LO stands for: \N{0}\LO=\LO, \N{1}\LO=\NLO, \N{2}\LO=\NNLO and so forth. Equation~\eqref{Eq:first} implies that if $\qt>0$ the infrared (IR) divergences that appear in the computation of $\rd\sigma^{F}_{\N{n}\LO}|_{\qt>0}$ are those already present in $\rd\sigma^{F+\jets}_{\N{n-1}\LO}$.
Therefore, provided that the IR singularities involved in $\rd\sigma^{F+\jets}_{\N{n-1}\LO}$ can be handled and cancelled with the available subtraction methods at \N{n-1}\LO, the only remaining singularities at  \N{n}\LO are associated with the limit $\qt \rightarrow 0$ and we treat them with the $\qt$ subtraction method. Since the small-$\qt$ behaviour of the transverse momentum distribution is well known through the resummation program~\cite{qTRes:program} of logarithmically-enhanced contributions to transverse-momentum distributions, we can (in principle) exploit this knowledge to construct the necessary \N{n}\LO counterterms (\CT) to subtract the remaining singularity at $\qt=0$, thereby promoting the $\qt$ subtraction method proposed in Refs.~\cite{Catani:2007vq} to \N{n}\LO. 

The generic form of the $\qt$ subtraction method~\cite{Catani:2007vq} for the \N{n}\LO cross section is
\begin{align}
  \frac{\rd\sigma^{F}_{\N{n}\LO}}{\rd\qt^2\rd M^2\rd Y}
  &=
  \cH^F_{\N{n}\LO} \otimes
  \frac{\rd\sigma^{F}_{\LO}}{\rd\qt^2\rd M^2\rd Y} +
  \left[ 
    \frac{\rd\sigma^{F+\jets}_{\N{n-1}\LO}}{\rd\qt^2\rd M^2\rd Y} -
    \frac{\rd\sigma^{F\,\CT}_{\N{n}\LO}(\qt)}{\rd\qt^2\rd M^2\rd Y}
  \right] \;,
  \label{eq:master}
\end{align}
where the symbol ``$\otimes$'' denotes convolutions over the momentum fractions and the flavour indices of the incoming partons and is explicitly defined as
\begin{align}
  \cG(\dots) \otimes \frac{\rd\sigma^{F}}{\rd\obs} 
  &\equiv
  \int_0^1\rd\xi_1 \int_0^1\rd\xi_2 \;
  \int_0^1\rd z_1  \int_0^1 \rd z_2 \; 
  \nonumber\\&\quad
  \times 
  \frac{\rd\hat{\sigma}^{F}_{ab}(\xi_1 z_1 p_1, \xi_2 z_2 p_2)}{\rd\obs} \;
  \cG_{ab \gets cd}(\ldots; z_1, z_2) \;
  f_{c/h_1}(\xi_1,\muf) \; f_{d/h_2}(\xi_2,\muf) \;.
  \nn\\&=
  \int_0^1\rd x_1 \int_0^1\rd x_2 \;
  \int_{x_1}^1 \frac{\rd z_1}{z_1} \int_{x_2}^1 \frac{\rd z_2}{z_2} \; 
  \nonumber\\&\quad
  \times 
  \frac{\rd\hat{\sigma}^{F}_{ab}(x_1 p_1, x_2 p_2)}{\rd\obs} \;
  \cG_{ab \gets cd}(\ldots; z_1, z_2) \;
  f_{c/h_1}\left(\frac{x_1}{z_1},\muf\right) \; f_{d/h_2}\left(\frac{x_2}{z_2},\muf\right) \;.
\end{align}
The counterterm $\rd\sigma^{F\,\CT}_{\N{n}\LO}$ constitutes the contribution to the \N{n}\LO cross section which cancels the divergences of $\rd\sigma^{F+\jets}_{\N{n-1}\LO}$ in the limit $\qt \rightarrow 0$ and renders the term in square brackets finite for all values of $\qt$. 
The $n$-th order counterterm can be written as
\begin{align}
  \frac{\rd\sigma^{F\,\CT}_{\N{n}\LO}(\qt)}{\rd\qt^2\rd M^2\rd Y}
  &=
  \Sigma^F_{\N{n}\LO}(\qt) \otimes
  \frac{\rd\sigma^{F}_{\LO}}{\rd M^2\rd Y} \;,
  \label{eq:CT}
\end{align}
where we note that the dependence of the function $\Sigma^{F}_{\N{n}\LO}(\qt)$ on the transverse momentum $\qt$ is \emph{not} kinematically related to the Born-level process.

The functions $\Sigma^{F}_{\N{n}\LO}(\qt)$ and $\cH^{F}_{\N{n}\LO}$ correspond to the $n$-th order truncation of the perturbative series in $\as$ of the functions
\begin{align}
  \Sigma^F_{c\bar{c} \gets ab}(\qt; z_1,z_2) 
  &=
  \sum_{n=1}^\infty \asOpi^n \; 
  \Sigma^{F;(n)}_{c\bar{c} \gets ab}(\qt; z_1,z_2) 
  \;,
  \label{eq:sigexpansion}
  \\
  \cH^F_{c\bar{c} \gets ab}(z_1,z_2) 
  &=
  \delta_{c\,a}\delta_{\bar{c}\,b} \, \delta(1-z_1) \, \delta(1-z_2) +
  \sum_{n=1}^\infty \asOpi^n \; 
  \cH^{F;(n)}_{c\bar{c} \gets ab}(z_1,z_2) 
  \;,
  \label{Hstexpand}
\end{align}
where the labels $a$ and $b$ stand for the partonic channels of the \N{n}\LO correction that are mapped to that the Born cross section. 
The function $ \Sigma^{F}(\qt)$ embodies all the terms of the form $\log(\qt^2/M^2)$ that are divergent in the limit $\qt \rightarrow 0$ and reproduces the logarithmically singular behaviour of  $\rd\sigma^{F+\jets}$ in the small-$\qt$ limit. 
Terms proportional to $\delta(\qt^{2})$ as well as IR finite terms are absorbed in the perturbative factor $\mathcal{H}^{F}$. The hard coefficient function $\mathcal{H}^{F}_{\N{n}\LO}$ thus encodes all the IR finite terms of the $n$-loop contributions.

According to the transverse momentum resummation formula~\cite{Bozzi:2005wk} and using the Fourier transformation between the conjugate variables $\qt$ and the impact parameter $b$, the perturbative hard function $\cH^{F}$ and the corresponding counterterm are obtained by the fixed-order truncation of the identity 
\begin{align}
  \Bigl( \Sigma^F(\qt)  + \cH^F\,\delta(\qt^2) \Bigr) &\otimes 
  \frac{\rd\sigma^{F}_{\LO}}{\rd M^2\rd Y} 
  =
  \frac{1}{s}\int_0^\infty\rd b\;\frac{b}{2}\;J_0(b\qt)\;
  \rd\hat{\sigma}^{F}_{\LO;c\bar{c}}(x_1 p_1, x_2 p_2) \; 
  S_c(M,b) \; 
  \nn\\&\quad\times
\int_{x_{1}}^1 \frac{\rd z_1}{z_{1}} \int_{x_{2}}^1  \frac{\rd z_2}{z_{2}} 
  \left[ H^{F} C_1 C_2 \right]_{c\bar{c};ab} \;
  f_{a/h_1}\left(\frac{x_1}{z_1},\frac{b_0^2}{b^2}\right) \; f_{b/h_2}\left(\frac{x_2}{z_2},\frac{b_0^2}{b^2}\right) \;.
  \label{reslean}
\end{align}
where  $b_0=2 \mathrm{e}^{-\gamma_E}$ ($\gamma_E=0.5772\ldots$  is the Euler--Mascheroni constant).

The large logarithmic corrections are exponentiated in the Sudakov form factor $S_c(M,b)$ of the quark  ($c=q, {\bar q}$) or of the gluon ($c=g$), which has the following  expression:
\begin{align}
  S_c(M,b) &= 
  \exp \left\{ - \int_{b_0^2/b^2}^{M^2} \frac{\rd q^2}{q^2} 
  \left[ A_c(\as(q^2)) \;\ln \frac{M^2}{q^2} + B_c(\as(q^2)) \right] \right\} 
  \;,
  \label{formfact}
\end{align}
where the functions $A$ and $B$ permit a perturbative expansion in $\as$:
\begin{align}
  A_c(\as) &= 
  \sum_{n=1}^\infty \asOpi^n A_c^{(n)} \;,
  &
  B_c(\as) &= 
  \sum_{n=1}^\infty \asOpi^n B_c^{(n)} \;.
  \label{eq:abexp}
\end{align}
Explicit expressions for the coefficients $A_g^{(n)}$ and $B_g^{(n)}$ that are relevant for Higgs production are collected in Appendix~\ref{app:fixed-order} up to $n=3$.
In particular, we also give the $B_g^{(3)}$ coefficient in the hard resummation scheme as needed to evaluate Eq.~\eqref{reslean} for $F=H$ at \N3\LO.

The analytical form of the function $\Sigma^{F;(3)}$ in Eq.~\eqref{eq:sigexpansion} can be obtained by expanding Eq.~\eqref{reslean} to the corresponding matching order. The full analytical formula for $\Sigma^{F}$ is resummation scheme independent order by order in the strong coupling constant. Therefore, the logarithmic singular behaviour for $\Sigma^{F}$ at $\qt\rightarrow 0$ at 
each given order in $\as$ does not depend on the 
resummation scheme, and can be validated against the behaviour of the fixed-order results at small $\qt$. To fully 
account for the logarithmically enhanced terms at a given order requires a sufficient depth in the resummation accuracy prior to 
its fixed-order expansion in Eq.~\eqref{eq:sigexpansion}. Specifically, the LO Higgs boson $\qt$ distribution receives singular contributions from up to NLL (next-to-leading-logarithm) resummation~\cite{Catani:1988vd,Kauffman:1991cx}, the NLO Higgs boson $\qt$ distribution requires the expansion of NNLL resummation~\cite{deFlorian:2001zd,deFlorian:2000pr,Becher:2012yn,Neill:2015roa}, and the NNLO Higgs boson $\qt$ distribution has been recently validated against the singular contributions from N$^3$LL resummation~\cite{Chen:2018pzu,Bizon:2018foh}. 

The structure of the symbolic factor denoted by $\left[ H^{F} C_1 C_2 \right]_{c\bar{c};a b}$ in Eq.~\eqref{reslean}, depends on the initial-state channel of the Born subprocess and is explained in detail in Refs.~\cite{Catani:2010pd,Catani:2013tia}. Here we limit ourselves to the case in which the final state system $F$ is composed of a single Higgs boson, $F\equiv H$, in which case,
\begin{align}
  \left[ H^{H} C_1 C_2 \right]_{gg;ab}
  = 
  H_{g}^{H}\left(\as(M^2)\right) \; \Bigl[ \; &
    C_{g \,a}\left(z_1;\as(b_0^2/b^2)\right) \; 
    C_{g \,b}\left(z_2;\as(b_0^2/b^2)\right) 
    \nn\\
    {}+{} & 
    G_{g \,a}\left(z_1;\as(b_0^2/b^2)\right) \; 
    G_{g \,b}\left(z_2;\as(b_0^2/b^2)\right)
  \;\Bigr]
  \;,
  \label{whathlean}
\end{align}
where $H_{g}^{H}$ is the hard--virtual function and respectively $C_{g \,a}$ and $G_{g \,a}$ the gluonic helicity-preserving and helicity-flipping hard--collinear coefficient functions.

The gluonic hard--collinear coefficient function $C_{g \,a}(z;\as)$ ($a=q,{\bar q},g$) has the following perturbative expansion
\begin{align}
  C_{g \,a}(z;\as) &= 
  \delta_{g \,a} \; \delta(1-z) + 
  \sum_{n=1}^\infty \asOpi^n C_{g\, a}^{(n)}(z) \;.
  \label{cgexp} 
\end{align}
In contrast, the perturbative expansion of the helicity flip hard--collinear coefficient function $G_{ga}$, which is specific to gluon-initiated processes, starts only at ${\cal O}(\as)$, and can be expanded as~\cite{Catani:2010pd,Catani:2013tia}
\begin{align}
  G_{g \,a}(z;\as) &=
  \sum_{n=1}^\infty \asOpi^n G_{g \,a}^{(n)}(z) \;.
  \label{gfexp}
\end{align}
The IR finite contribution of the $n$-loop correction terms to the Born subprocess is contained in the hard--virtual function (which does not depend on $z_1$ or $z_2$),
\begin{align}
  H_g^{H}(\as) &= 
  1+ \sum_{n=1}^\infty \asOpi^n H_g^{H \,;(n)} \;.
  \label{hexp}
\end{align} 

Using Eqs.~\eqref{reslean} and \eqref{whathlean}, then, after integration over $b$ and dropping the renormalisation group predictable terms that are produced by evolving $\as$ to a common scale (i.e. setting $\muF=\muR=M$), we obtain the resummation scheme independent
\begin{align}
  \cH^H_{gg\gets ab}(z_1,z_2; \muf=\mur=M)
  &\equiv 
  H_g^H(\as) \Big[ 
    C_{g \,a}(z_1;\as) \; C_{g \,b}(z_2;\as) +
    G_{g \,a}(z_1;\as) \; G_{g \,b}(z_2;\as)
  \Big] \; .
  \label{HCCGG}
\end{align}
Note that in the literature, it is often the rapidity-integrated variant $\cH^H_{gg\gets ab}(z)$ that is quoted which is related to $\cH^H_{gg\gets ab}(z_1,z_2)$ via the convolution
\begin{align}
  \cH^H_{gg\gets ab}(z)
  &\equiv
  \int_0^1\rd z_1 \; \int_0^1\rd z_2 \; \delta(z-z_1 z_2) \; \cH^H_{gg\gets ab}(z_1,z_2) \;.
\end{align}

The $\cH^H$ function in Eq.~\eqref{HCCGG} can be expanded perturbatively without approximation to any order in the strong coupling constant $\as$. 
Inserting the expansions of the hard functions into Eq.~\eqref{HCCGG}, then,
\begin{align}
  \cH^{H;(1)}_{gg\gets ab}(z_1,z_2; \muf=\mur=M)
  &=
  \delta_{g\,a} \,\delta_{g\,b} 
  \,\delta(1-z_1) \,\delta(1-z_2) 
  \,H^{H;(1)}_g 
  \nn\\&\quad
  +\delta_{g\,a} \,\delta(1-z_1) \,C^{(1)}_{g\,b}(z_2)
  +\delta_{g\,b} \,\delta(1-z_2) \,C^{(1)}_{g\,a}(z_1) \; , 
  \label{H1}
  \\
  \cH^{H;(2)}_{gg\gets ab}(z_1,z_2; \muf=\mur=M)
  &=
  \delta_{g\,a} \,\delta_{g\,b}
  \,\delta(1-z_1) \,\delta(1-z_2) 
  \,H^{H;(2)}_g
  \nn\\&\quad
  +\delta_{g\,a} \,\delta(1-z_1) \,C^{(2)}_{g\,b}(z_2)
  +\delta_{g\,b} \,\delta(1-z_2) \,C^{(2)}_{g\,a}(z_1)
  \nn\\&\quad
  +H^{H;(1)}_g\left(
  \delta_{g\,a} \,\delta(1-z_1) \,C^{(1)}_{g\,b}(z_2) +
  \delta_{g\,b} \,\delta(1-z_2) \,C^{(1)}_{g\,a}(z_1)
  \right)
  \nn\\&\quad
  +C^{(1)}_{g\,a}(z_1) \, C^{(1)}_{g\,b}(z_2)
  +G^{(1)}_{g\,a}(z_1) \, G^{(1)}_{g\,b}(z_2) \; .
  \label{H2}
\end{align}
Explicit expressions for the known fixed-order coefficients are collected in Appendix~\ref{app:fixed-order}. 

The new third-order contribution is given by
\begin{align}
  \cH^{H;(3)}_{gg\gets ab}(z_1,z_2; \muf=\mur=M)
  &=
  \delta_{g\,a} \,\delta_{g\,b} \,\delta(1-z_1)\,\delta(1-z_2)\,H^{H;(3)}_g
  \nn\\&\quad
  +\delta_{g\,a} \,\delta(1-z_1)\,C^{(3)}_{g\,b}(z_2)
  +\delta_{g\,b} \,\delta(1-z_2)\,C^{(3)}_{g\,a}(z_1)
  \nn\\&\quad
  +G^{(1)}_{g\,a}(z_1) G^{(2)}_{g\,b}(z_2)
  +G^{(2)}_{g\,a}(z_1) G^{(1)}_{g\,b}(z_2)
  \nn\\&\quad
  +H^{H;(1)}_g\left(
  \delta_{g\,a} \,\delta(1-z_1) \,C^{(2)}_{g\,b}(z_2) +
  \delta_{g\,b} \,\delta(1-z_2) \,C^{(2)}_{g\,a}(z_1)
  \right) 
  \nn\\&\quad
  +H^{H;(2)}_g\left(
  \delta_{g\,a} \,\delta(1-z_1) \,C^{(1)}_{g\,b}(z_2) +
  \delta_{g\,b} \,\delta(1-z_2) \,C^{(1)}_{g\,a}(z_1)
  \right)
  \nn\\&\quad
  + H^{H;(1)}_g  C^{(1)}_{g\,a}(z_1) \, C^{(1)}_{g\,b}(z_2) 
  + H^{H;(1)}_g  G^{(1)}_{g\,a}(z_1) \, G^{(1)}_{g\,b}(z_2) 
  \nn\\&\quad
  + C^{(1)}_{g\,a}(z_1) \, C^{(2)}_{g\,b}(z_2)
  + C^{(2)}_{g\,a}(z_1) \, C^{(1)}_{g\,b}(z_2) \;.
  \label{H3}
\end{align}
The second-order helicity-flip functions $G^{(2)}_{g\,a}(z)$, the third-order collinear functions $C^{(3)}_{g\,a}(z)$ and the third-order hard--virtual coefficient $H^{H;(3)}_g$ are only known in parts or not at all, thereby presenting an obstacle to applying the $\qt$ subtraction formalism at \N3\LO.
Nevertheless, within the $\qt$ subtraction  formalism, all these resummation coefficients can be inferred for any hard scattering process whose corresponding total cross section is known at \N3\LO. 
This point is discussed in detail in Sect.~\ref{sec:numforCN3}.

Although the hard--virtual coefficient $H^{H;(3)}_{g}$ is currently not known in analytical form, parts of it can be inferred from known results in threshold resummation.
This relies on the knowledge of the general structure of $H^{F}_{c}$ (to all orders), which relates $H^{F;(n)}_{c}$ to the finite part of the $n$-loop virtual Matrix Element~\cite{Catani:2013tia}.
To this end, we split $H^{H;(3)}_{g}$ into two pieces,
\begin{align}
  H^{H;(3)}_g  
  &\equiv  
  \widetilde{H}^{H;(3)}_{g} + \big[H^{H;(3)}_{g}\big]_{(\delta^{\qt}_{(2)})} \;,
  \label{H3deltaqT2}
\end{align}
where $\widetilde{H}^{H;(3)}_{g}$ can be computed using the corresponding hard--virtual factor $C^{\mathrm{th}(3)}_{gg\to H}$~\cite{Catani:2014uta} from threshold resummation (in the large-$m_t$ limit) and the exponential equation that relates hard--virtual coefficients in threshold- and $\qt$-resummation (Eq.~(81) of Ref.~\cite{Catani:2013tia}). We find,
\begin{align}
  \widetilde{H}^{H;(3)}_{g}
  &=
  C_{A}^{3} \biggl(
    -\frac{15649 \zeta_{3}}{432}
    -\frac{121 \pi ^2 \zeta_{3}}{432}
    +\frac{3\zeta_{3}^2}{2}
    +\frac{869 \zeta_{ 5}}{144}
    +\frac{215131}{5184}
    +\frac{16151 \pi ^2}{7776}
    -\frac{961\pi ^4}{15552}
    \nn\\&\qquad
    +\frac{\pi ^6}{810}
    +\frac{105}{32} \zeta_{6}
  \biggr)
  + C_{A}^{2} \biggl(
    \frac{605 \zeta_{ 3}}{72}
    +\frac{55 \pi ^2 \zeta_{ 3}}{36}
    +\frac{737 \pi ^2}{432}
    +\frac{167 \pi ^4}{432}
    +\frac{\pi^6}{72}
  \biggr) 
  \nn\\&\quad
  + C_{A} \biggl( 
    \frac{19 \pi ^2  L_{t}}{48}
    -\frac{55 \pi ^2 \zeta _{3}}{8}
    -\frac{\pi ^6}{480}
    +\frac{133 \pi ^4}{72}
    +\frac{11399\pi ^2}{864}  
    +\frac{63}{32} \zeta_{6}
  \biggr)
  \nn\\&\quad
  + N_{f}^{2} \biggl( 
    \frac{43 C_{A} \zeta_{ 3}}{108}
    -\frac{19 \pi ^4 C_{A}}{3240}
    -\frac{133 \pi ^2 C_{A}}{1944}
    +\frac{2515C_{A}}{1728}
    -\frac{7 C_{F} \zeta_{ 3}}{6}
    \nn\\&\qquad
    +\frac{4481 C_{F}}{2592}
    -\frac{\pi ^4C_{F}}{3240}
    -\frac{23 \pi ^2 C_{F}}{432} 
  \biggr)
  \nn\\&\quad
  + N_{f} \biggl(
    \frac{101 C_{A}^2 \zeta_{5}}{72}
    -\frac{97}{216} \pi ^2 C_{A}^2 \zeta_{3}
    +\frac{29 C_{A}^2 \zeta_{3}}{8}
    +\frac{1849 \pi ^4 C_{A}^2}{38880}
    -\frac{35 \pi ^2 C_{A}^2}{243}
    -\frac{98059C_{A}^2}{5184}
    \nn\\&\qquad
    +\frac{5 C_{A} C_{F} \zeta _{5}}{2}
    +\frac{13 C_{A} C_{F} \zeta_{3}}{2}
    +\frac{1}{2} \pi ^2 C_{A} C_{F} \zeta _{3}
    -\frac{63991 C_{A} C_{F}}{5184}
    +\frac{11 \pi ^4C_{A} C_{F}}{6480}
    \nn\\&\qquad
    -\frac{71}{216} \pi ^2 C_{A} C_{F}
    +\frac{1}{9} \pi ^2 C_{A}L_{t}
    -\frac{5}{36} \pi ^2 C_{A} \zeta _{3}
    -\frac{55 C_{A} \zeta_{ 3}}{36}
    -\frac{5 \pi ^4C_{A}}{54}
    -\frac{1409 \pi ^2 C_{A}}{864}
    \nn\\&\qquad
    -5 C_{F}^2 \zeta_{ 5}
    +\frac{37 C_{F}^2 \zeta_{3}}{12}
    +\frac{19 C_{F}^2}{18}
  \biggr) \;,
  \label{H3approx}
\end{align}
with $L_t=\ln(M^2/m_t^2)$ and $\zeta_n$ denoting the Riemann zeta-function for integer values $n$ ($\zeta_2=\pi^2/6$, $\zeta_3=1.202\dots$, $\zeta_4=\pi^4/90$).
Note that we neglect all the third-order terms in the exponent of Eq.~(81) in Ref.~\cite{Catani:2013tia}, considering the entire $\mathcal{O}(\as^{3})$ correction (in the exponent) as unknown. However, the full top-mass dependence of $H^{H;(3)}_g$ is already fully embodied in $\widetilde{H}^{H;(3)}_{g}$.
The currently unknown $\big[H^{H;(3)}_{g}\big]_{(\delta^{\qt}_{(2)})}$ represents a \emph{single} coefficient (of \emph{soft} origin) belonging to the finite part of the structure of the IR singularities contained in the third-order virtual amplitude of the corresponding partonic subprocess $gg \to H$. 

As a consequence, the only missing ingredients to $\cH^{H;(3)}$ are the functions $G^{(2)}_{g\,a}(z)$, $C^{(3)}_{g\,a}(z)$ and $ \big[H^{H;(3)}_{g}\big]_{(\delta^{\qt}_{(2)})}$. 
The details on their numerical extraction will be discussed in the following section.

\section{The Higgs boson total cross section at \texorpdfstring{N${}^\text{3}$LO}{N3LO}}
\label{sec:numforCN3}

We start this section by reviewing some properties of the hard-scattering function $\cH_{c{\bar c} \gets ab}^{F}$. 
This function is resummation-scheme independent, but it depends on the specific hard-scattering subprocess $c + {\bar c} \to F$. 
The coefficients $\cH_{c{\bar c} \gets ab}^{F;(n)}$ of the perturbative expansion in Eq.~\eqref{Hstexpand} can be determined by performing a perturbative calculation of the $\qt$ distribution in the limit $\qt \to 0$. 
In the right-hand side of Eq.~\eqref{reslean}, the function $\cH^{F}$ controls the strict perturbative normalization of the corresponding total cross section (i.e.\ the integral of the total $\qt$ distribution). 
This unitarity-related property can be exploited to determine the coefficients $\cH_{c{\bar c} \gets ab}^{F ;(n)}$ from the perturbative calculation of the inclusive cross section.  
In particular, the integral of the full $\qt$ spectrum in Eq.~\eqref{eq:master} must reproduce the inclusive cross section $\sigma^{F\,\tot}$,
\begin{align}
  \sigma^{F\,\tot}_{\N{n}\LO}
  &=
  \int_0^\infty\rd\qt^2 \; \frac{\rd\sigma^{F}_{\N{n}\LO}}{\rd\qt^2} \;, 
  &
  \frac{\rd\sigma^{F}_{\N{n}\LO}}{\rd\qt^2}
  &\equiv
  \int\rd M^2 \, \rd Y \;
  \frac{\rd\sigma^{F}_{\N{n}\LO}}{\rd\qt^2\rd M^2\rd Y} \;.
  \label{restotp}
\end{align}
Since the hard-scattering function $\cH_{c\bar{c}\gets ab}^{F}$ is accompanied by $\delta(\qt^{2})$, we evaluate the $\qt$ spectrum on right-hand side of Eq.~\eqref{eq:master} according to the following decomposition \cite{Bozzi:2005wk}
\begin{align}
  \sigma^{F\,\tot}_{\N{n}\LO}
  &=
  \cH^F_{\N{n}\LO} \otimes \sigma^F_{\LO} + 
  \int_0^\infty\rd\qt^2 \; \frac{\rd\sigma^{F\,\fin}_{\N{n}\LO}}{\rd\qt^2} \;, 
  \label{sigtotrel} 
\end{align}
where $\rd{\sigma}^{F\,\fin}$ is directly related to the quantity in square brackets in the right-hand side of Eq.~\eqref{eq:master}
\begin{align}
  \frac{\rd\sigma^{F\,\fin}_{\N{n}\LO}}{\rd\qt^2}
  &\equiv
  \left[
    \frac{\rd\sigma^{F+\jets}_{\N{n-1}\LO}}{\rd\qt^2} -
    \frac{\rd\sigma^{F\,\CT}_{\N{n}\LO}}{\rd\qt^2} 
  \right]
  \; .
  \label{sigfin}
\end{align}

The relation in Eq.~\eqref{sigtotrel} is valid order-by-order in QCD perturbation theory~\cite{Bozzi:2005wk}. If the perturbative coefficients of the fixed-order expansion of $\sigma^{F\,\tot}$, $\cH^F$ and $\rd \sigma^{F\,\fin}/\rd \qt^2$ are all known, the relation~\eqref{sigtotrel} has to be regarded as an identity, which can be explicitly checked. Since the fixed-order truncation of $\rd \sigma^{F\,\fin}/\rd \qt^2$ is free of any contribution proportional to $\delta(\qt^2)$, its NLO contribution  does not contain the coefficient $\cH^{F;(1)}$, and so forth. 
Therefore, $\cH^{F;(3)}$ can be isolated from the  the  \N3\LO  term in Eq.~\eqref{sigtotrel}:
\begin{align}
  \left[
    \sigma^{F\,\tot}_{\N3\LO} - 
    \sigma^{F\,\tot}_{\N2\LO}
  \right] -
  \int_0^\infty\rd\qt^2 \left[
    \frac{\rd\sigma^{F\,\fin}_{\N3\LO}}{\rd\qt^2} -
    \frac{\rd\sigma^{F\,\fin}_{\N2\LO}}{\rd\qt^2}
  \right]
  &=
  \asOpi^3 \cH^{F;(3)} \otimes \sigma^F_{\LO} \;,
  \label{h3nnlo}
\end{align}
where $\as=\as(\muR^2)$.

If all the components on the left-hand side of Eq.~\eqref{h3nnlo} are known analytically (as it was the case at NNLO in Refs.~\cite{Catani:2011kr,Catani:2012qa}), the function $\cH^{F}$ can be extracted exactly in analytical form. At NLO the extraction of the function  $\cH^{F;(1)}$ is straightforward for Drell--Yan and Higgs boson production. The function $\cH^{F;(2)}$ at NNLO (for Higgs ($F=H$) boson production~\cite{Catani:2011kr} and Drell--Yan ($F=DY$)~\cite{Catani:2012qa}) can be obtained with a dedicated analytical computation using the analogue of Eq.~\eqref{h3nnlo} at NNLO. Since the transverse momentum distributions for $H$+jet and DY+jet at NNLO are not known analytically, Eq.~\eqref{h3nnlo} can be used only numerically to compute $\cH^{F;(3)}$. 

As was elaborated on at the end of the previous section, the general structure of the coefficient $\cH^{F;(3)}$ is not known in analytic form for any hard-scattering process. Nonetheless, within the $\qt$ subtraction  formalism, $\cH^{F;(3)}$ can be reliably approximated for any hard-scattering process whose corresponding total cross section is known at \N3\LO. 
As identified in Eqs.~\eqref{H3} and \eqref{H3deltaqT2}, the only missing ingredients to $\cH^{F;(3)}$ are the functions $G^{(2)}_{g\,a}(z)$, $C^{(3)}_{g\,a}(z)$ and $ \bigl[H^{H;(3)}_{g}\bigr]_{(\delta^{\qt}_{(2)})}$. 
Their contribution to Eq.~\eqref{h3nnlo} can be approximated as follows:
\begin{align}
  &
  \left[ H_g^{H;(3)} \right]_{\delta^{\qt}_{(2)}} 
  \delta_{g\,a} \, \delta(1-z_1) \, \delta_{g\,b} \, \delta(1-z_2) 
  \nn\\&
  + C^{(3)}_{g\,a}(z_1) \, \delta_{g\,b} \, \delta(1-z_2) 
  + \delta_{g\,a} \, \delta(1-z_1) \, C^{(3)}_{g\,b}(z_2) 
  + G^{(2)}_{g\,a}(z_1) \, G^{(1)}_{g\,b}(z_2) 
  + G^{(1)}_{g\,a}(z_1) \, G^{(2)}_{g\,b}(z_2) 
  \nn\\&\qquad\approx
  C_{N3} \; \delta_{g\,a} \, \delta(1-z_1) \, \delta_{g\,b} \, \delta(1-z_2) \;,
  \label{CN3eq}
\end{align}
where the third-order coefficient $C_{N3}$ embodies the numerical extraction of the hard--virtual coefficient $\big[H^{H;(3)}_{g}\big]_{(\delta^{\qt}_{(2)})}$  \textit{plus} the approximation of the $z_i$-dependent functions by a numerical constant proportional to $\delta(1-z_i)$. The resulting coefficient $\big[H^{H;(3)}_{g}\big]_{(\delta^{\qt}_{(2)})}$ is exact since $C_{N3}$ is proportional to $\delta(1-z)$ (or equivalently $\delta(1-z_1)\delta(1-z_2)$). In other words, the approximation that is made in Eq.~\eqref{CN3eq} is related only to the functions $G^{(2)}_{g\,a}(z)$ and $C^{(3)}_{g\,a}(z)
$, whose functional dependence on the variable $z$ goes beyond terms proportional to $\delta(1-z)$, and which involves not only gluon-to-gluon transitions ($a=g$), but also contributions from 
other parton species ($a=q,\bar q$). The latter are not explicitly distinguished in the above approximation, which fully attributes their numerical contribution to the gluon-induced processes. 

The method outlined in Eq.~\eqref{CN3eq} to approximate the unknown terms in the hard--virtual function $\cH^H_{gg\gets ab}$ numerically is not new. It was first used in Ref.~\cite{Bozzi:2005wk} in order to compute the second order function $\cH^{H;(2)}_{gg\gets ab}$ numerically at NNLO, providing a reasonable estimate of the exact result to better than $1\%$ accuracy. Notice that Eq.~\eqref{CN3eq} ensures that one recovers the total cross section (at \N3\LO in this case) with no approximation. After integration over the transverse momentum $\qt$, Eq.~\eqref{restotp} provides the same total integral (numerically in this case)  as in the fully analytical case. Even more, for IR-safe observables (at fixed order) where the \textit{back-to-back} kinematical configuration ($\qt=0$) is located at a single phase space point (e.g.\ the $\qt$ distribution, the angular separation $\Delta \varphi _{\gamma\gamma}$ between the two photons for a Higgs boson decaying into diphotons, etc.), the fixed order result is also exact, i.e.\ the integral of the analytical unknown terms in Eq.~\eqref{CN3eq} (which all have $\qt=0$) is located in one single point of the exclusive differential distributions. 

The previous considerations about the approximation underpinning Eq.~\eqref{CN3eq} were regarding the total cross section or differential distributions in which the Born-like configurations belong to one single phase space point.
In order to quantify the quality of the approximation proposed in Eq.~\eqref{CN3eq} at the differential level when the Born differential cross section populates the entire differential range, we perform a detailed numerical study of the Higgs boson rapidity $Y \equiv y_{H}$ distribution in Sec.~\ref{sec:NNLOrap} at NNLO. Anticipating these results,  we find that in the rapidity range $0\leq y_{H}\leq 4$ the approximated NNLO result differs by less than $0.2\%$ from the exact NNLO Higgs boson rapidity distribution.

\subsection{Implementation and setup of the numerical calculations}
\label{sec:numsetup}

To extract the value of $C_{N3}$, we first introduce the numerical tools and the calculational setup in this section. We use the same setup for the inclusive and differential predictions presented in Sections~\ref{sec:numCN3}, \ref{sec:NNLOrap}, \ref{sec:N3LOrap} and \ref{sec:results}.

We consider Higgs boson production in proton--proton collisions at a centre-of-mass energy of $\sqrt{s}=13$~TeV. In our computation, we set the Higgs boson mass to $M\equiv M_H= 125$~GeV and the vacuum expectation value to $v=246.2$~GeV. 
The Born process is initiated via gluon--gluon fusion mediated through a top-quark loop, which can be integrated out in the large-$m_t$ limit ($m_{t}\rightarrow \infty$).
In this limit, the production of the Higgs boson is described through an effective gluon-gluon-Higgs boson vertex~\cite{Heft}.
The mass of the top quark is taken as $m_t = 173.2$~GeV, which enters in the contributions that have a residual $m_t$ dependence (e.g.\ Eqs.~\eqref{H2g} and \eqref{H3approx} and effective vertex coefficient corrections at \N3\LO). 
With the top quark loop replaced by an effective vertex, we consider a five-flavour scheme QCD with all light quarks being massless. We use the central set of the \verb|PDF4LHC15| PDFs~\cite{nnpdf} as implemented in the \texttt{LHAPDF} framework~\cite{Buckley:2014ana} and the associated strong coupling constant with $\as(M_Z)=0.118$. Note that we systematically employ the same order in the PDFs (in particular the set \verb|PDF4LHC15_nnlo_mc|) for the LO, NLO, NNLO and \N3\LO results presented in this paper. The central factorization and renormalization scale is chosen as $\mu \equiv \muR = \muF =  M_H / 2$. The theoretical uncertainty is estimated by varying the default scale choice independently for $\muR$ and $\muF$ by factors of $\{1/2,2\}$ while omitting combinations with $\muR/\muF = 4$ or $1/4$, resulting in the common seven-point variation of scale combinations. 

As stated in Sec.~\ref{sec:numforCN3} and in Ref.~\cite{Catani:2007vq}, the computation of the total cross section or differential distributions with the  $\qt$ subtraction formalism can be separated into two main parts by inserting Eq.~\eqref{sigfin} into Eq.~\eqref{sigtotrel}:
\begin{align}
  \sigma^{F\,\tot}_{\N{n}\LO}
  &=
  \left[
    \cH^F_{\N{n}\LO} \otimes \sigma^F_{\LO}
    - \int_0^\infty\rd\qt^2 \; \frac{\rd\sigma^{F\,\CT}_{\N{n}\LO}}{\rd\qt^2} 
  \right]
  + \int_0^\infty\rd\qt^2 \; \frac{\rd\sigma^{F+\jets}_{\N{n-1}\LO}}{\rd\qt^2} \;.
  \label{sigtotnumerical} 
\end{align}
The contribution $\rd{\sigma}^{F+\jets}$ in Eq.~\eqref{sigtotnumerical} is computed with the parton-level event generator \texttt{NNLOJET} which provides the necessary infrastructure for the antenna subtraction method up to NNLO~\cite{Antenna:method}. 
Processes at NNLO with the structure of $\rd{\sigma}^{F+\jets}$ implemented in \texttt{NNLOJET} are: $F=H$~\cite{Chen:2016zka}, $F=\gamma^*,~Z$~\cite{Ridder:2015dxa,Gehrmann-DeRidder:2016jns} and $F=W^{\pm}$~\cite{Gehrmann-DeRidder:2017mvr}. In this paper we focus on Higgs production $F=H$, where the relevant matrix elements in 
 \texttt{NNLOJET} are: ($H+1$)-parton production at two loops~\cite{Gehrmann:2011aa}, ($H+2$)-parton production at one loop~\cite{Dixon:2009uk,Badger:2009hw,Badger:2009vh} and ($H+3$)-parton 
 production at tree-level~\cite{DelDuca:2004wt,Dixon:2004za,Badger:2004ty}. The subtraction formalism that we are applying to Higgs boson production could be easily extended  to $Z$ and $W^{\pm}$ production~\cite{leaninprep}.

The terms in square brackets in Eq.~\eqref{sigtotnumerical}  for $F=H$ are encoded in a new Monte Carlo generator \texttt{HN3LO}~\cite{leaninprepHN3LO} up to the third order in the strong coupling constant. After expanding Eq.~\eqref{reslean} to this order, several non-trivial convolutions emerge and we briefly document the corresponding formulae implemented in \texttt{HN3LO} in Appendix~\ref{app:Convos}. All our results up to the NNLO level are in full agreement with the Monte Carlo generator \texttt{HNNLO}~\cite{Catani:2007vq} at the per mille level of accuracy. On the left-hand side of Eq.~\eqref{h3nnlo}, the Higgs boson total cross sections at NNLO ($\sigma^{H\,\tot}_{\NNLO}$) and \N3\LO  ($\sigma^{H\,\tot}_{\N{3}\LO}$) are also required. 
We use the analytical coefficient function for the total Higgs boson cross section that was recently calculated in Ref.~\cite{Mistlberger:2018etf} and which is available within the public program \texttt{ihixs 2}~\cite{Dulat:2018rbf}.
This program is further used to compute any of the analytical total cross-section ingredients required to extract the coefficient $C_{N3}$.

The numerical computation of the integral of the difference $\rd{\sigma}_{\NNLO}^{F+\jets}-\rd{\sigma}_{\N3\LO}^{F\,\CT}$ in Eq.~\eqref{sigfin}, although finite, requires the introduction of a suitable technical lower bound or $\qtcut$, since both terms in this difference are logarithmically divergent at $\qtcut \to 0$. 
 This technical cut introduces systematic uncertainties to both $\rd{\sigma}_{\NNLO}^{F+\jets}$ and $\rd{\sigma}_{\N3\LO}^{\,\CT}$. Once cancellations between the terms on the right-hand side of Eq.~\eqref{sigtotnumerical} take place, the numerically calculated total cross sections and differential distributions have to be $\qtcut$ independent (within the statistical errors) over some range of $\qtcut$. At the lower end of this range, 
numerical instabilities in $\rd{\sigma}_{\NNLO}^{F+\jets}$ (arising from the large dynamical range in this calculation) 
will limit the accuracy of the result, while at the 
higher end of the range, missing non-logarithmic terms in  $\rd{\sigma}_{\N3\LO}^{F\,\CT}$ will start to 
become significant. 
The numerical stability of $\rd{\sigma}_{\NNLO}^{F+\jets}$ at small $\qt$ using \texttt{NNLOJET} has been systematically validated for Higgs boson production (with $\qtcut=0.7$~GeV in Ref.~\cite{Chen:2018pzu}) and Drell--Yan production (with $\qtcut=2$~GeV in Ref.~\cite{Bizon:2018foh}) at the LHC. In Sections~\ref{sec:numCN3}, \ref{sec:NNLOrap}, \ref{sec:N3LOrap} and \ref{sec:results}, we document numerical results obtained with the $\qt$ subtraction formalism using $\qtcut=(2\pm 1)$~GeV.

\subsection{The numerical extraction of \texorpdfstring{$C_{N3}$}{CN3}}
\label{sec:numCN3}

In the following, we describe the numerical results regarding the extraction of the $C_{N3}$ coefficient and the corresponding \N3\LO total cross section. 

\begin{figure}[tb]
\centering
\includegraphics[width=.6\linewidth]{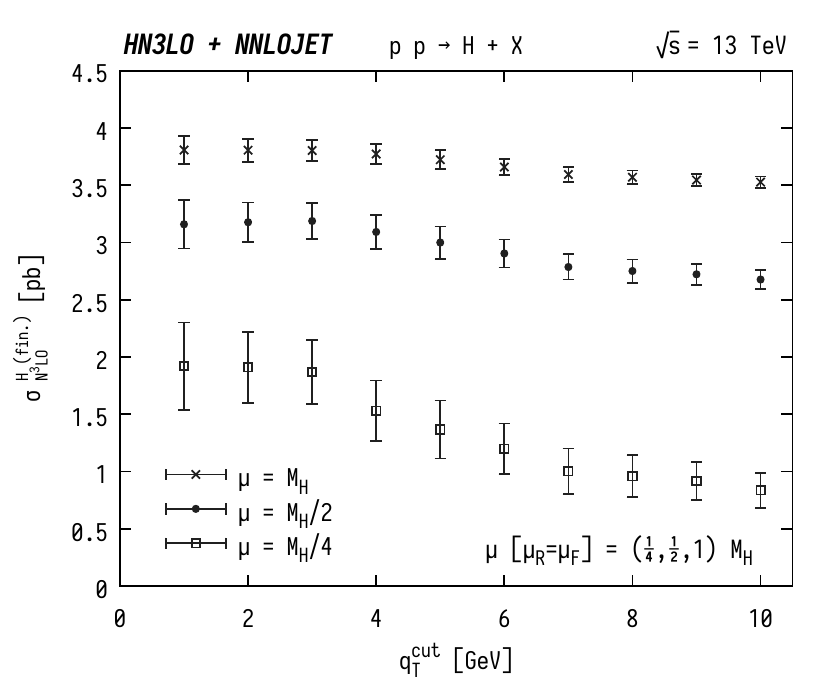}
\caption{\label{fig:finiten3lo}{The $\qt$ integrated finite contribution to the cross section of Eq.~\eqref{sigfin} at \N3\LO-only (i.e.\ $\N3\LO-\NNLO$) between $\qtcut$ and $\infty$, for three different scales ($\mu=\muR=\muF$).}}
\end{figure}

In Fig.~\ref{fig:finiten3lo} we display the $\sigma^{H\,\fin}_{\N3\LO}$ at \N3\LO-only \emph{coefficient} as a function of the $\qtcut$, i.e.,  the difference $\sigma^{H\,\fin}_{\N3\LO} - \sigma^{H\,\fin}_{\NNLO}$. 
The error bars denote the numerical integration errors from  \texttt{NNLOJET}. Since the figure displays cumulant cross sections as function of the lower integration boundary, the central values and errors are fully correlated among the points.
Using Eq.~\eqref{H3} with Eq.~\eqref{h3nnlo} and the value of the resulting integral $ \sigma^{H\,\fin}(\qtcut=1~\mathrm{GeV})$ in Fig.~\ref{fig:finiten3lo}, it is possible to obtain the $\qt$-integrated cross section of the unknown terms on the left-hand side of Eq.~\eqref{CN3eq} and consequently extract $C_{N3}$. 

The behaviour of $\sigma^{H\,\fin}_{\N3\LO}$ as a function of $\qtcut$ is shown in Fig.~\ref{fig:finiten3lo} and gives an estimate of the systematical uncertainty corresponding to the use of this technical cut which turns out to be at the \textit{per mille} level in the domain $\qtcut=(2\pm 1)$~GeV for the total Higgs boson cross section at \N3\LO. More specifically, variations of the $\qtcut$ parameter from $\qtcut=1$~GeV to 3~GeV produce variations in the central value of the 
\N3\LO contribution to $\sigma^{H\,\fin}$ cross section of less than $0.1\%$ for the scales $\mu=M_{H}$ and $\mu=M_{H}/2$, and variations of the order of $0.3\%$ for $\mu=M_{H}/4$. These variations are considerably smaller than the numerical integration error at fixed $\qtcut$. 

\begin{figure}
\centering
\includegraphics[width=.6\linewidth]{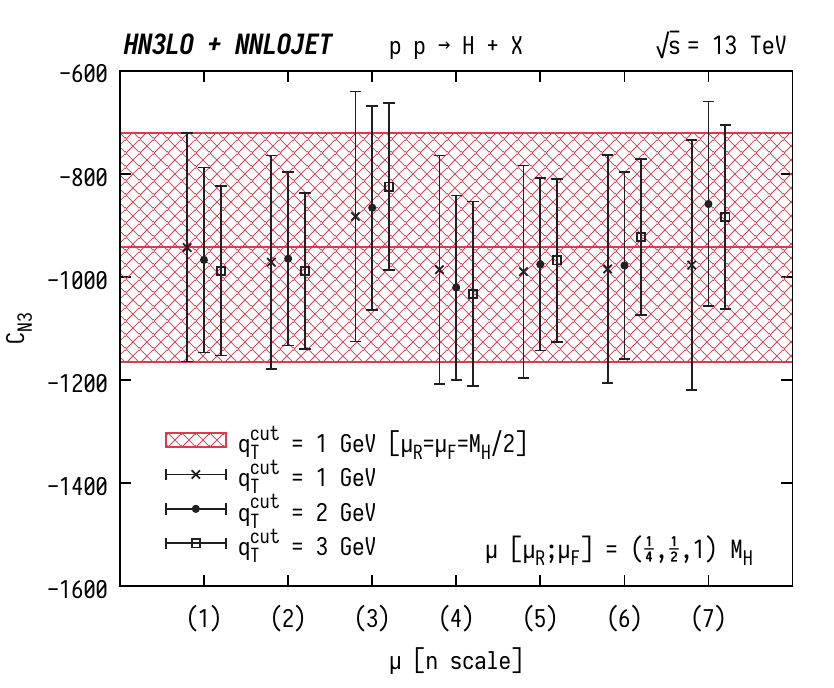}
\caption{\label{fig:CN3qT}{The numerically extracted $C_{N3}$ coefficient (for three different values of $\qtcut$) as a function of the combination of scales, as enumerated in Table~\ref{Table:CN3values}.
The error bars for each particular $C_{N3}$ point are obtained propagating the statistical uncertainties of the different terms involved in the computation. The red band corresponds to our best estimation for $C_{N3}$ obtained with the central scale $\mu=M_{H}/2$ at $\qtcut=1$~GeV, as detailed in the text.}}
\end{figure}

\begin{table}
\centering
\renewcommand{\arraystretch}{1.1}
\begin{tabular}{ |c|c||c|c|c|c|c|c| }

\hline

\multirow{5}{*}{ } \!\!\!\!n &
\multicolumn{1}{ |p{30mm}| }{ $\big[\tilde{\muR},\tilde{\muF} \big]$ $\times M_{H}$} &
\multicolumn{1}{ |p{30mm}| }{$C_{N3}$ {\scriptsize ( $\qtcut=1$~GeV) }} &
\multicolumn{1}{ |p{30mm}| }{$C_{N3}$ {\scriptsize ( $\qtcut=2$~GeV) }} &
\multicolumn{1}{ |p{30mm}| }{$C_{N3}$ {\scriptsize ( $\qtcut=3$~GeV) }} \\
\hline

\multirow{1}{*}{(1)} 

& $\big[1/2,1/2 \big]$ 

&$ \mathbf{ -943~\pm 222}$ 

& $ -967~\pm 179$

& $-988~\pm 164$

 \\ 

\multirow{1}{*}{(2)} 

& $\big[1,1 \big]$

& $-971~\pm 207$ 

& $ -965~\pm 168$

& $-989~\pm 151$

 \\ 

\multirow{1}{*}{(3)} 

& $\big[1/4,1/4 \big]$

& $ -883~\pm 243$ 

& $ -866 ~\pm 198$

& $-850~\pm 162$

\\ 

\multirow{1}{*}{(4)} 

& $\big[1/2,1 \big]$

& $ -986~\pm 222$ 

& $ -1021~\pm 179$

& $-1033~\pm 179$

 \\

\multirow{1}{*}{(5)} 

& $\big[1,1/2 \big]$

& $ -990~\pm 206$ 

& $ -976~\pm 167$

& $ -968~\pm 158$

 \\
 
\multirow{1}{*}{(6)} 

& $ \big[1/2,1/4 \big]$

& $ -985~\pm 221$ 

& $ -978~\pm 181$

& $-923~\pm 152$

\\

\multirow{1}{*}{(7)} 

& $\big[1/4,1/2 \big]$

& $ -977~\pm 243$ 

& $ -859~\pm 199$

& $ -883~\pm 179$
\\

\hline

\end{tabular}

\caption{\label{Table:CN3values}
{Extracted values of the $C_{N3}$ coefficients as a function of the $\qtcut$ as shown in Fig.~\ref{fig:CN3qT} for each scale choice. In bold typeface the $C_{N3}$ coefficient (for the case $\qtcut=$1~GeV) which constitutes our best estimation. The uncertainty for each one of the $C_{N3}$ coefficients is determined with the customary propagations of the uncertainties. The first column is used to label each particular scale choice used in Fig.~\ref{fig:CN3qT}.
}
}
\renewcommand{\arraystretch}{1}
\end{table}

In Table~\ref{Table:CN3values} and Figure~\ref{fig:CN3qT},
we collect the values of $C_{N3}$ extracted for all seven combinations of scale 
choices and three different values of $\qtcut$. 
We note that the central value of each $C_{N3}$ is independent of the scale (within the uncertainties), in complete agreement with Eq.~\eqref{H3}. This scale independence of $C_{N3}$ is unrelated to the ansatz of Eq.~\eqref{CN3eq}: the terms in the right-hand side of Eq.~\eqref{H3} are all scale independent and the relation between $C_{N3}$ and $\widetilde{H}^{H;(3)}_{g}$ is defined through Eqs.~\eqref{H3}, \eqref{H3deltaqT2} and \eqref{CN3eq}. The uncertainties shown in Fig.~\ref{fig:CN3qT} are determined using conventional error propagation and are almost entirely dominated by the size of the statistical errors of the \N3\LO $\sigma_{H}^{\fin}$ cross section shown in Fig.~\ref{fig:finiten3lo}.

Since the resulting cross sections at different scale values are statistically correlated, we propose as our estimation for the $C_{N3}$ coefficient the value obtained for  $\qtcut=1$~GeV   at the central scale $\muF=\muR=M_{H}/2$, $C_{N3}=-943 \pm 222$, which is indicated in bold typeface in Table~\ref{Table:CN3values}.
The solid red central line in Fig.~\ref{fig:CN3qT}, and the associated red band are obtained using this single value. 

\begin{figure}
\centering
\includegraphics[width=.6\linewidth]{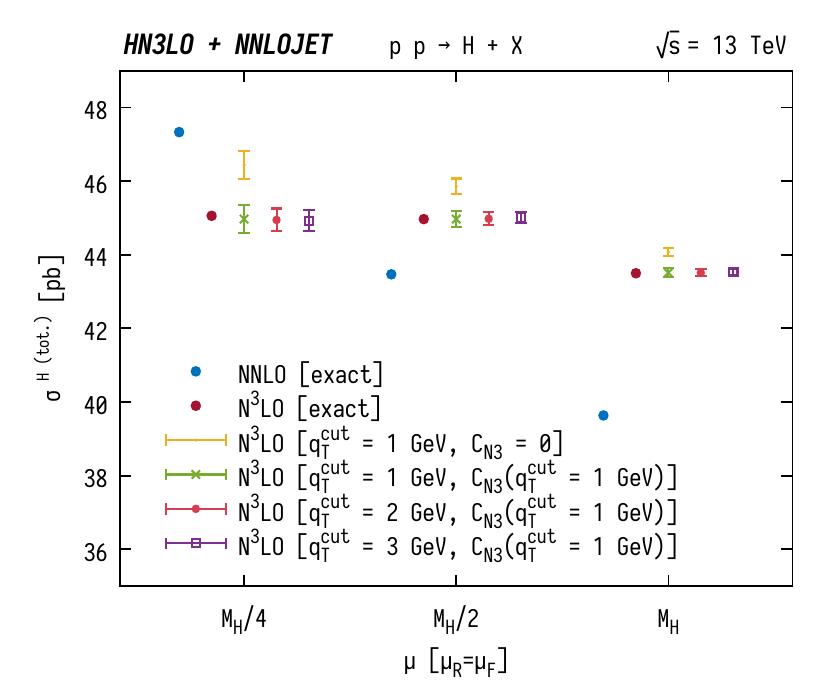}
\caption{\label{fig:totXsecN3LO}{Total cross section of Higgs boson production, $\sigma^{H\,\tot}_{\N{3}\LO}$, as obtained by the $\qt$ subtraction formalism, compared with the corresponding analytical $\sigma^{H\,\tot}_{\N3\LO}$ of Ref.~\cite{Mistlberger:2018etf} (dark red dots). Green crosses with error bar denote the $\qt$ subtraction prediction for $\qtcut=1$~GeV, red dots with error bar represents  $\sigma^{H\,\tot}_{\N3\LO}$ using $\qtcut=2$~GeV, and purple square dots with error bar having $\qtcut=3$~GeV. Whereas the $\qtcut$ is changed (from 1 to 3~GeV) the coefficient $C_{N3}$ is always fixed to be the value extracted in Fig.~\ref{fig:CN3qT} for $\qtcut=1$~GeV. The $\qt$ subtraction prediction at \N3\LO with the $C_{N3}$ numerical coefficient fixed to zero (using $\qtcut=1$~GeV) is shown using yellow dots with error bar. The NNLO analytical Higgs boson cross section ($\sigma^{{H\,\tot}}_{\N3\LO}$) is represented by blue dots. All the cross sections are shown for three different scales: $\mu \equiv \muR = \muF = \{1/4,1/2,1 \} M_H$ and horizontally displaced for better visibility. The uncertainty bars in the $\qt$ subtraction predictions are calculated with the customary propagation of statistical uncertainties.}}
\end{figure}

\begin{table}
\centering
\renewcommand{\arraystretch}{1.1}
\resizebox{\linewidth}{!}{%
\begin{tabular}{ |c||c|c|c|c|c|c| }

\hline

\multirow{5}{*}{ } $\sigma^{H\,\tot}$~(pb)
&\multicolumn{1}{ |c| }{Exact} &
\multicolumn{1}{ |p{30mm}| }{\rm $\qt$~ subtraction \newline  {\scriptsize ($\qtcut=1$~GeV)}} &
\multicolumn{1}{ |p{30mm}| }{\rm $\qt$~ subtraction \newline  {\scriptsize ($\qtcut=2$~GeV) }} &
\multicolumn{1}{ |p{30mm}| }{\rm $\qt$~ subtraction \newline  {\scriptsize ($\qtcut=3$~GeV)}} &
\multicolumn{1}{ |p{30mm}| }{\rm $\qt$~ subtraction \newline  {\scriptsize ($C_{N3}=0$)}}\\
\hline

\multirow{1}{*}{~~~\N3\LO {\scriptsize$\big[\mu=M_{H}/2\big]$}} 

& $44.97$  

& $44.97~\pm 0.21$ 

& $44.98~\pm 0.17$ 

& $45.01~\pm 0.15$ 

& $45.86~\pm 0.21$ \\ 

\multirow{1}{*}{\N3\LO  {\scriptsize$\big[\mu=M_{H}\big]$} } 

& $43.50$

& $43.51~\pm 0.12$ 

& $43.51~\pm 0.10$ 

& $43.53~\pm 0.09$ 

&$44.08~\pm 0.12$ \\ 

\multirow{1}{*}{~~~\N3\LO  {\scriptsize$\big[\mu=M_{H}/4\big]$} } 

& $45.06$

& $44.97~\pm 0.38$ 

& $44.95~\pm 0.31$ 

& $44.92~\pm 0.28$ 

&$46.44~\pm 0.38$ \\ 

\hline

\multirow{1}{*}{~~~NNLO  {\scriptsize$\big[\mu=M_{H}/2\big]$} } 

& $43.47$

& $43.46~\pm 0.02$

& $43.46~\pm 0.02$

& $43.46~\pm 0.02$ 

& $43.46~\pm 0.02$  \\

\multirow{1}{*}{NNLO  {\scriptsize$\big[\mu=M_{H}\big]$} } 

& $39.64$

& $39.62~\pm 0.02$ 

& $39.62~\pm 0.02$ 

& $39.62~\pm 0.02$ 

& $39.62~\pm 0.02$ \\

\multirow{1}{*}{~~~NNLO  {\scriptsize$\big[\mu=M_{H}/4\big]$} } 

& $47.33$

& $47.33~\pm 0.02$  

& $47.33~\pm 0.02$

& $47.33~\pm 0.02$  

& $47.33~\pm 0.02$ \\

\hline

\end{tabular}
}
\caption{\label{Table:totXsec}
{The total cross section for Higgs boson production $\sigma^{H\,\tot}$ at the LHC ($\sqrt{s}=13$~TeV). Results for NNLO and \N3\LO cross sections for three different scales $\mu=M_{H}/2$ (central scale), $\mu=M_{H}$ and $\mu=M_{H}/4$. The column \lq{}\lq{}Exact\rq{}\rq{} contains the results of Ref.~\cite{Mistlberger:2018etf} computed with the numerical code of Ref.~\cite{Dulat:2018rbf} as detailed in the text. The results with the $\qt$ subtraction method are obtained using three different values of $\qtcut$ (1,2 and 3~GeV), and their uncertainties are calculated with the customary propagation of statistical errors. The last column shows $ \sigma^{H\,\tot}$ obtained with the $\qt$ subtraction method and using $C_{N3}=0$ at \N3\LO. The values of $ \sigma^{H\,\tot}$ reported in this Table are shown in Fig.~\ref{fig:totXsecN3LO}. The NNLO cross sections computed with the $\qt$ subtraction method are obtained using $\qtcut=1$~GeV, i.e.\ the variation of this parameter in the \N3\LO cross section is considered at \N3\LO-only.
}
}
\renewcommand{\arraystretch}{1}
\end{table}

The numerically extracted $C_{N3}$ coefficient allows the total cross section to be computed at \N3\LO using 
the $\qt$ subtraction method, which serves as a closure test of the approach and the approximations used,  and allows the impact of uncertainties associated with the numerical evaluation of the ingredients to be quantified. 
In Fig.~\ref{fig:totXsecN3LO} we compare the fully analytical \N3\LO Higgs boson total cross section~\cite{Mistlberger:2018etf} (dark red dot) and our estimation (red dot with error bar) for three central scales, using $\qtcut=2$~GeV. The yellow dots with error bar represent our best approximation without the use of the $C_{N3}$ coefficient (i.e.\ $C_{N3}=0$), that can be considered as the prediction of the $\qt$ subtraction method in the case in which the total cross section is unknown (e.g.\ for Drell--Yan at \N3\LO). The uncertainty bars in the $\qt$ subtraction prediction correspond to the statistical errors of the numerical computations and are mainly due to the finite contribution in Eq.~\eqref{sigfin} at \N3\LO-only. The green crosses and purple squares correspond to our \N3\LO prediction using $\qtcut=1$~GeV and 3~GeV respectively. Notice that the $\qtcut$ variation is performed at \N3\LO-only, while the NNLO cross section is evaluated at fixed $\qtcut$ parameter.
The NNLO cross section is also shown in Fig.~\ref{fig:totXsecN3LO} (blue dots) in order to put the size of the \N3\LO corrections in relation to the previous perturbative order. The total cross sections shown in Fig.~\ref{fig:totXsecN3LO} are reported in Table~\ref{Table:totXsec}.

\section{The rapidity distribution of the Higgs boson}
\label{sec:rapgen}

In this section we use the $C_{N3}$ coefficient (extracted in Sec.~\ref{sec:numCN3}) to produce differential predictions at \N3\LO. In particular, we present differential results for the rapidity distribution of the Higgs boson. In Sec.~\ref{sec:NNLOrap}  we first estimate at NNLO the uncertainties introduced in the rapidity distribution by the procedure proposed in Eq.~\eqref{CN3eq}. In Sec.~\ref{sec:N3LOrap} we present the rapidity distribution at \N3\LO with the
 estimation of the uncertainties associated to the variation of the $\qtcut$ and $C_{N3}$ parameters.

\subsection{The NNLO rapidity distribution}
\label{sec:NNLOrap}

In this section we aim to quantify the uncertainty in the approximation used in Eq.~\eqref{CN3eq}. This approximation was first proposed in Ref.~\cite{Bozzi:2005wk} for Higgs production at NNLO. Since all the ingredients of the $\qt$ subtraction formalism at NNLO are known in analytical form~\cite{Catani:2011kr}, it is possible to quantify the difference induced by the approximation compared to the exact result. This analysis further allows to assess the potential impact of the approximation that could be present at \N3\LO in Sec.~\ref{sec:N3LOrap} and \ref{sec:results} below. For this quantitative study we consider the collinear functions $C^{(1)}_{g\,a}$ and the hard--virtual factor $H^{H;(1)}_g$ in Eq.~\eqref{H2} as known. The collinear functions $C^{(2)}_{g\,a}$ and the first order helicity-flip functions $G^{(1)}_{g\,a}$ are regarded as unknown. The hard--virtual factor $H^{H;(2)}_g$ is divided in two contributions in analogy to Eq.~\eqref{H3deltaqT2}
\begin{align}
  H^{H;(2)}_g  
  &\equiv  
  \widetilde{H}^{H;(2)}_{g} + \big[H^{H;(2)}_{g}\big]_{(\delta^{\qt}_{(1)})} \;,
\end{align}
where $\big[H^{H;(2)}_{g}\big]_{(\delta^{\qt}_{(1)})}$ is considered as unknown for the present NNLO study. 

\begin{figure}[t]
\begin{minipage}{.48\linewidth}
  \centering
  \includegraphics[width=\linewidth]{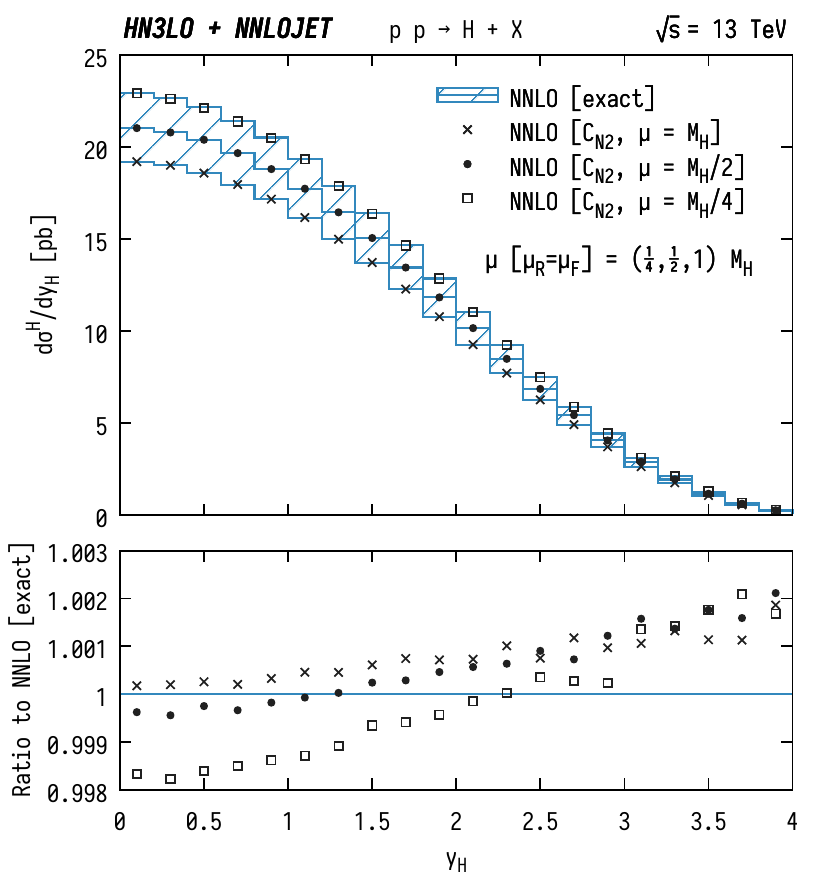}
  \\
  (a)
\end{minipage}
\hfill
\begin{minipage}{.48\linewidth}
  \centering
  \includegraphics[width=\linewidth]{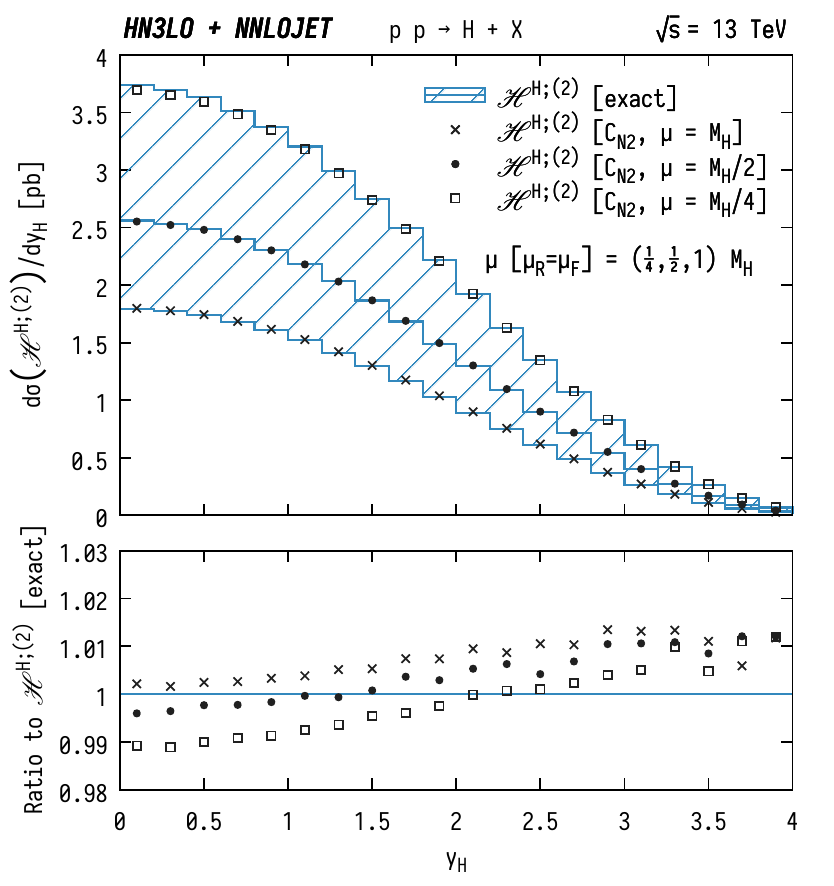}
  \\
  (b)
\end{minipage}
\caption{\label{fig:yHCN}{Comparison of the rapidity distribution between the exact result at NNLO (blue hatched) and an evaluation using the $C_{N2}$ numerical coefficient (cross, dot and square points). 
We perform the comparison both at the level of (a)~the full NNLO cross section and (b)~for the coefficient function $\cH^{H;(2)}$.
The lower panels show the ratio to the exact result. 
For this particular example at NNLO, we employ the three-point scale variation: $\mu=\muR=\muF=\{M_{H}/4,M_{H}/2,M_{H}\}$. 
}}
\end{figure}

These so-called \emph{unknown functions} (for this exercise) which depend on the variables $z_i$ in Eq.~\eqref{H2} are approximated with a single numerical coefficient $C_{N2}$ proportional to $\delta(1-z_1)\delta(1-z_2)$ (the $C_{N2}$ here was labeled as $C_{N}$ in Ref.~\cite{Bozzi:2005wk}) in direct analogy to Eq.~\eqref{CN3eq}:
\begin{align}
  &
  \left[ H_g^{H;(2)} \right]_{\delta^{\qt}_{(1)}} 
  \delta_{g\,a} \, \delta(1-z_1) \, \delta_{g\,b} \, \delta(1-z_2) 
  \nn\\&
  + C^{(2)}_{g\,a}(z_1) \, \delta_{g\,b} \, \delta(1-z_2) 
  + \delta_{g\,a} \, \delta(1-z_1) \, C^{(2)}_{g\,b}(z_2) 
  + G^{(1)}_{g\,a}(z_1) \, G^{(1)}_{g\,b}(z_2) 
  \nn\\&\qquad\approx
  C_{N2} \; \delta_{g\,a} \, \delta(1-z_1) \, \delta_{g\,b} \, \delta(1-z_2) \;,
  \label{CNeq}
\end{align}
In Fig.~\ref{fig:yHCN}(a) we show the rapidity distribution of the Higgs boson at NNLO computed with the exact $\qt$ subtraction (blue hatched band) and the NNLO prediction using the $C_{N2}$ coefficient (dot, cross and square points). For this particular example at NNLO, we employ the three-point scale variation: $\mu=\muR=\muF=\{M_{H}/4,M_{H}/2,M_{H}\}$. Repeating the analysis performed for Table~\ref{Table:CN3values} and Fig.~\ref{fig:CN3qT}, we obtain: $C_{N2}=28\pm 1$. The numerical value of the $C_{N2}$ parameter corresponds to a specific $\widetilde{H}^{H;(2)}_{g}$ hard coefficient:
\begin{align}
\label{Ht2g}
\widetilde{H}^{H;(2)}_{g}&=\frac{11399}{144}+\frac{19}{8} L_{t}-\frac{1189}{144} N_{f}+\frac{2}{3} N_{f} L_{t}+\frac{83}{6} \pi^{2} -\frac{5}{18} \pi^{2} N_{f} + \frac{13}{16} \pi^{4} - \frac{165}{4} \zeta_{3} + \frac{5}{6} N_{f} \zeta_{3}\;\;,
\end{align}
which is obtained with the same method that was used to arrive at Eq.~\eqref{H3approx}.
Using this $C_{N2}$ parameter we can produce differential predictions which are obtained \textit{mimicking} the strategy that we intend to apply at \N3\LO. 

In the lower panel of Fig.~\ref{fig:yHCN}(a) we show the ratio to the exact NNLO result, i.e.\ we present the ratio for each scale. As expected, the approximation presents its best behaviour at central rapidity and the deviation from the exact results is at \textit{per mille} level throughout the considered rapidity range of $|y_{H}|\leq 4$.

The study shown in Fig.~\ref{fig:yHCN}(a) validates the quality of our method for the total rapidity distribution of the Higgs boson at NNLO. One could argue that a more stringent check would involve only the quantities involved in the approximation, i.e., the rapidity distribution of the second-order coefficient functions $\cH^{H;(2)}$. 

In Fig.~\ref{fig:yHCN}(b) we compare the rapidity distribution for $\cH^{H;(2)}_{\rm exact}$ (defined in Eq.~\eqref{H2}) with the approximated $y_{H}$ distribution of the coefficient $\cH^{H;(2)}_{\rm C_{N2}}$, defined in Eq.~\eqref{CNeq}. The function $\cH^{H;(2)}_{\rm C_{N2}}$ approximates the exact $\cH^{H;(2)}_{\rm exact}$  within a precision of $2\%$, demonstrating the accuracy of the proposed method even at the level of individual coefficients. This directly implies that the contribution of the hard--virtual factor $H^{H;(2)}_g$ is more important than the rapidity--dependent functions $G^{(1)}_{g\,a}(z)$ and $C^{(2)}_{g\,a}(z)$ across the whole rapidity range.

We performed at NNLO variations of the $\qtcut$ value between 0.1~GeV and 3~GeV, and the NNLO cross sections (and differential distributions) present deviations within a range of size $0.26\%$  (the largest deviation is always observed for the scale choice $\mu=M_{H}/4$).  We consider $\qtcut=1$ GeV enough to proceed at NNLO (and as our reference value), as we can understand from Table \ref{Table:totXsec} at NNLO.

Summarizing, we have presented in this subsection a validation at NNLO of the approximation used at \N3LO. We have performed two kinds of tests: i) a check over the observable and ii) a validation at the level of the coefficients involved in the approximation. While case ii) establishes the quality of the approach regarding the approximated particular quantities, case i) evaluates the precision of the approximation at the level of the observable which is the decisive and strongest test.

\subsection{Numerical stability of the \texorpdfstring{N${}^\text{3}$LO}{N3LO} rapidity distribution}
\label{sec:N3LOrap}

\begin{figure}[tbh]
\centering
\includegraphics[width=.6\linewidth]{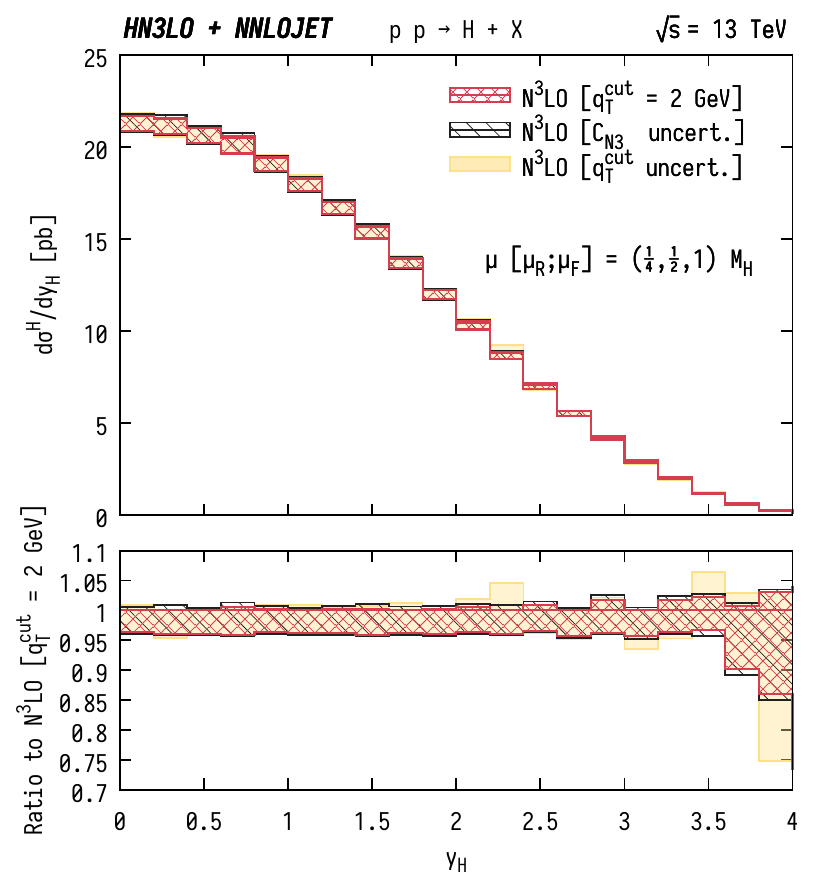}
\caption{\label{fig:yHN3LOonly}{Rapidity distribution of the Higgs boson as computed using the $\qt$ subtraction formalism at \N3\LO. All bands include the seven-point scale variation as detailed in Table~\ref{Table:CN3values}. The red band constitutes our result with $\qtcut=2$~GeV using the central value for the $C_{N3}$ coefficient ($C_{N3}=-943$). The pale yellow band is obtained as the envelope between the prediction at $\qtcut=1$~GeV and $2$~GeV using $C_{N3}=-943$. The black band is computed at fixed $\qtcut=2$~GeV  taking the two extremal values of the C$_{N3}$ coefficient according to the uncertainty ($C_{N3}=-943 \pm 222$), and performing seven-point scale variation as described in the text.
}}
\end{figure}

In this section, we quantify the numerical stability (as well as the involved intrinsic uncertainties) of the Higgs boson rapidity distribution at \N3\LO concerning the $\qtcut$ and $C_{N3}$ parameters and the statistical uncertainties introduced by $\rd\sigma^{H\,\fin}/\rd y_H$ at \N3\LO-only.

In Fig.~\ref{fig:yHN3LOonly} we show the rapidity distribution at \N3\LO obtained with the $\qt$ subtraction method using the $C_{N3}$ coefficient determined in Sec.~\ref{sec:numCN3} ($C_{N3}=-943 \pm 222$). The NNLO prediction is always computed with $\qtcut=1$~GeV. 
The red band in Fig.~\ref{fig:yHN3LOonly} shows the size of the seven-point scale variation for $\qtcut=2$~GeV. 

The pale yellow band is calculated as the envelope of the scale variation bands for two different values of $\qtcut$: 1~GeV and 2~GeV. Therefore, the pale yellow band in Fig.~\ref{fig:yHN3LOonly} can be taken as an estimate of the uncertainty due to the variation of the $\qtcut$ parameters at \N3\LO. In Fig.~\ref{fig:totXsecN3LO} (and Table~\ref{Table:totXsec}), we observed that the total cross section (for the three central scales) is rather stable as a function of the $\qtcut$ value. The variations of the \N3\LO cross sections were at the \textit{per mille} level of accuracy if we consider  $\qtcut=2 \pm 1$~GeV, which is far better than the associated statistical uncertainty (see Table~\ref{Table:totXsec}). The uncertainty estimate due to the $\qtcut$ variation performed in Fig.~\ref{fig:yHN3LOonly}, which is differential in the Higgs-boson rapidity, confirms the stability of the total cross section reported in Table~\ref{Table:totXsec}. 
The rapidity distribution is almost insensitive to the change in the $\qtcut$ parameter in the region where the bulk of the cross section is concentrated ($|y_{H}| \leq 3.6$). At large rapidities ($|y_{H}| \sim 4$), where the overall contribution to the total cross section is less than $0.5\%$, we found the largest deviations. Such deviations are mainly related to the numerical uncertainties from $\rd\sigma^{H\,\fin}/\rd y_H$ at  \N3\LO-only.
 
Finally, we consider the uncertainty introduced by the statistical errors of the $C_{N3}$ coefficient. The black band in Fig.~\ref{fig:yHN3LOonly} is obtained as the envelope of the seven-point scale variation at $\qtcut=2$~GeV now considering for each scale the two extremal $C_{N3}$ coefficients corresponding to its maximum and minimum statistical deviations: $C_{N3} =\{-1165,-721\}$. The envelope is therefore taken from a total of 14 rapidity distributions (two extremal predictions for each one of the seven scales). The net effect of this $C_{N3}$ variation result in an overall enlargement of the red band at $\qtcut=2$~GeV. Our final uncertainty estimate in the rapidity distribution of the Higgs boson at \N3\LO is computed as the envelope of three bands: seven-point scale variation only, combined with $\qtcut$ variation, and combined with $C_{N3}$ variation.

\subsection{The rapidity distribution of the Higgs boson at \texorpdfstring{N${}^\text{3}$LO}{N3LO}}
\label{sec:results}

In this section we present our predictions for the Higgs boson rapidity distributions at the LHC, applying the \N3\LO $\qt$ subtraction method presented in Sec.~\ref{sec:forma}. The setup of the calculation is summarised in Sec.~\ref{sec:numCN3}. 

\begin{figure}[tbh]
\centering
\includegraphics[width=.6\linewidth]{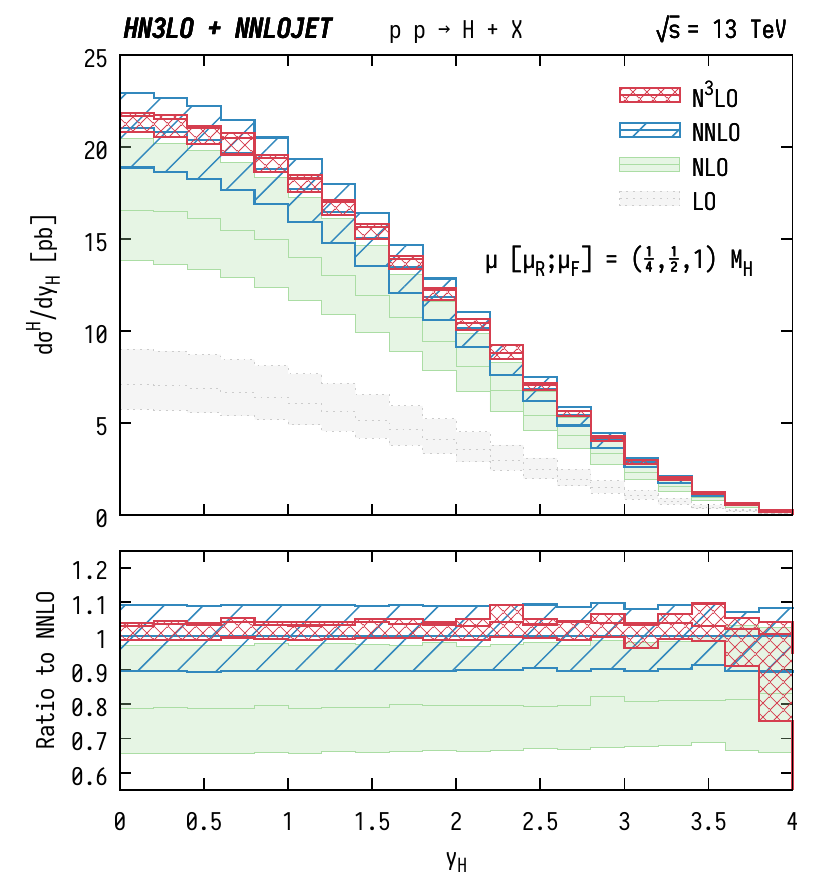}
\caption{\label{fig:yHN3LO}{Rapidity distribution of the Higgs boson computed using the $\qt$ subtraction formalism up to \N3\LO. The seven-point scale variation bands (as stated in Table~\ref{Table:CN3values}) of the LO, NLO, NNLO and \N3\LO($C_{N3}$) results are as follows: LO (pale grey fill), NLO (green fill), NNLO (blue hatched) and \N3\LO($C_{N3}$) (red cross-hatched). The central scale ($\mu=M_{H}/2$) at each perturbative order (except LO) is shown with solid lines. In the lower panel, the ratio to the NNLO prediction is shown. While the bands for the predictions at LO, NLO and NNLO are computed with the seven scales as detailed in the text, the \N3\LO($C_{N3}$) band is obtained after considering also the uncertainties due to the variation of the $\qtcut$ and the $C_{N3}$ coefficient in the 
\N3\LO-only contribution.
}}
\end{figure}

Figure~\ref{fig:yHN3LO} shows the rapidity distribution of the Higgs boson at LO (pale grey fill), NLO (green fill), NNLO (blue hatched) and \N3\LO (red cross-hatched). 
The central scale ($\mu=M_{H}/2$) is shown as a solid line while the bands correspond to the envelope of seven-point scale variation. 
At \N3\LO, the band additionally includes the uncertainties due to $\qtcut$ and $C_{N3}$ as described in Sec.~\ref{sec:N3LOrap}.
Going from LO to NNLO, the scale $\mu=M_{H}/2$ is always at the center of the respective scale variation band in Fig.~\ref{fig:yHN3LO}.  The central prediction at \N3\LO, on the other hand, almost coincides with the upper edge of the band, as was already observed for the total cross section~\cite{Anastasiou:2015ema,Mistlberger:2018etf}, see Table~\ref{Table:totXsec} and Fig.~\ref{fig:totXsecN3LO}.
Figures~\ref{fig:totXsecN3LO} and \ref{fig:yHN3LO} respectively show a substantial reduction in the size of the scale variation band at \N3\LO both in the total cross section and in differential distributions.

In the central rapidity region of $|y_{H}| \leq 3.6$, the impact of the \N3\LO corrections on the NNLO result is almost independent of $y_{H}$ with a flat $K$-factor about $1.034$ for the central scale choice. The combined theoretical uncertainty at \N3\LO is at most of $\pm 5 \%$ level with respect to the central scale choice. The uncertainty on the $y_H$ distribution is reduced by more than a factor of $1/2$ by going from NNLO to \N3\LO. The \N3\LO uncertainty band lies fully within the scale variation band at NNLO, exhibiting a stable perturbative behaviour. 
The only exception is the very high rapidity region, where the $\qtcut$ uncertainty becomes the dominant source for the size of the \N3\LO band as shown in Fig.~\ref{fig:yHN3LOonly}.

The \N3\LO corrections to the Higgs boson rapidity distribution have been investigated in Refs.~\cite{Dulat:2017prg,Dulat:2018bfe} employing a threshold expansion.
The first two leading terms in the threshold expansion were computed in Ref.~\cite{Dulat:2017prg}, which agrees well with our calculation for the rapidity region $y_H < 0.5$ despite different choices of PDFs and scale-variation prescriptions.
Both calculations display a considerable reduction of scale uncertainties going from NNLO to \N3\LO in this central rapidity region. 
For the rapidity region $y_H > 1$, however, larger differences are observed between the two calculations, where the results using the $\qt$ subtraction formalism generally yield smaller \N3\LO corrections (within the NNLO scale uncertainty band).
Most recently, the calculation of the threshold expansion including the first six terms was completed in Ref.~\cite{Dulat:2018bfe}, which exhibits a stabilisation of the perturbative series together with a reduction of scale uncertainties.
Comparing Fig.~\ref{fig:yHN3LO} with the results obtained in Ref.~\cite{Dulat:2018bfe}, we observe very good agreement between the two calculations.

\section{Conclusions and outlook}
\label{Sec:conclu}

In this paper we have performed a detailed study of Higgs boson production at the LHC using the $\qt$ subtraction formalism at \N3\LO. We systematically describe the $\qt$ subtraction formalism for a generic colourless and massive system $F(\{q_i\})$ produced at hadron colliders. Fully differential cross sections for this type of final state system are separated into $\delta(\qt)$ and $\qt\ne 0$ contributions. The contribution for $\qt\ne 0$ is calculated, 
using a phase space cut-off $\qtcut$, as the difference between $F(\{q_i\})+\jets$ production and $\qt$ counterterms. Specifically, we use the \texttt{NNLOJET} package to compute NNLO Higgs-plus-jet production and expand the Sudakov from factor in the hard resummation scheme to the matching order for the corresponding $\qt$ counterterms. The contribution at $\delta(\qt^{2})$ is further factorized into convolutions of the Sudakov form factor, the hard--virtual function, the helicity-flip coefficient function, the hard--collinear coefficient function as well as the PDFs (Sec.~\ref{sec:forma}). The factorization guarantees that all the process-dependent contributions proportional to a form factor are included in the hard--virtual function, which depends on both initial- and final-state particles. All other factorized contributions only depend on the initial states. Some of the factorized ingredients contributing at $\delta(\qt^{2})$ are not known analytically at \N3\LO for the moment. We collect all analytically available contributions and approximate the unknown pieces by a constant coefficient $C_{N3}$ which is scale- and process-independent (Sec.~\ref{sec:numforCN3}). Using the available inclusive total cross section for \N3\LO Higgs production and the known pieces from  the $\qt$ subtraction formalism, we numerically extract the value of $C_{N3}$. By comparing the numerical values for $C_{N3}$ using different scales and $\qtcut$ setups in the extraction, we conclude from mutually consistent results that $C_{N3}$ is independent of the scale choice with a value obtained for $\mu=M_H/2$ and $\qtcut = 1$~GeV of $C_{N3}=-943\pm 222$ (Sec.~\ref{sec:numCN3}).

As a proof-of-concept implementation of the $\qt$ subtraction method at \N3\LO,  we calculate the total cross section and rapidity distributions for Higgs boson production at LHC using a new Monte Carlo generator \texttt{HN3LO}~\cite{leaninprepHN3LO}. Using the extracted value of $C_{N3}$, we perform a closure test for the inclusive total cross section for three different scale choices and find excellent agreement with the exact results (from \texttt{ihixs 2}~\cite{Dulat:2018rbf}) at the $0.2\%$ level. For the differential rapidity distribution of the Higgs boson, we first study the systematic error from the $C_{N3}$ approximation by considering the NNLO calculation and introducing an approximate $C_{N2}$. The NNLO $y_H$ distribution exhibit per-mille level agreement between the $C_{N2}$ approximation and the exact result, supporting the reliability of the procedure. 
We calculate the $y_H$ distribution at \N3\LO employing a seven-point scale variation and carefully assess systematic errors arising form different $\qtcut$ and $C_{N3}$ values. Compared to the NNLO $y_H$ distributions, we observe a large reduction of theory uncertainties by more than $50\%$ at \N3\LO. The scale variation band at \N3\LO stays within the NNLO band with a flat $K$-factor of about $1.034$ in the central rapidity region ($|y_H|\leq3.6$). Both the systematic error analysis and the phenomenological predictions confirm that our calculations at \N3\LO using $\qt$ subtraction formalism are well under control. The approximation related to the $C_{N3}$ coefficient in our approach can be easily replaced by the full analytical results once available. 

With the upcoming larger data set and more accurate measurements of Higgs properties at the LHC, we prepare precise theoretical tools that could match the frontier accuracy of experimental results. More differential properties at \N3\LO involving the Higgs boson and its decay products can be studied using the same framework established in this paper. The current \N3\LO calculation, using the approximation of large top quark mass, attains a level of  precision that several 
other contributions will need to be taken into account for a full study of precision phenomenology~\cite{Anastasiou:2016cez}: finite top quark mass effects, heavy-light quark interference contributions  
and electroweak corrections.

\bigskip\noindent\textbf{Acknowledgements}

LC would like to thank Stefano Catani for very useful and valuable discussions. XC would like to thank Javier Mazzitelli and Hua Xing Zhu for inspiring discussions. We thank the University of Zurich S3IT and CSCS Lugano for providing the computational resources for this project. This research was supported in part by the UK Science and Technology Facilities Council, by the Swiss National Science Foundation (SNF) under contracts 200020-175595 and CRSII2-160814, by the Swiss National Supercomputing Centre (CSCS) under project ID UZH10, by the Research Executive Agency (REA) of the European Union under the ERC Advanced Grant MC@NNLO (340983).

\newpage
\section*{Appendix}
\begin{appendix}

\section{Fixed-order expressions}
\label{app:fixed-order}

The precise identification of the Sudakov form factor $S_c$, the hard--virtual function 
$H_g^{F=H}$ and the hard--collinear coefficient functions, $C_{g\, a}$ and $G_{g\, a}$ is not unique, and the resummation formula~\eqref{reslean} is invariant under ``resummation scheme''  transformations~\cite{Catani:2000vq}:
\begin{align}
  H_c^{F}(\as) &\to
  H_c^{F}(\as) \; \left[ \, h(\as) \, \right]^{-1} \;,
  \nn\\
  B_c(\as) &\to 
  B_c(\as) - \beta(\as) \;\frac{\rd\ln h(\as)}{\rd\ln \as} \;,
  \nn\\
  C_{ab}(\as,z) &\to
  C_{ab}(\as,z) \; \left[ \, h(\as) \, \right]^{1/2} \;,
  \nn\\
  G_{ab}(\as,z) &\to
  G_{ab}(\as,z) \;\left[ \, h(\as) \, \right]^{1/2} \;. 
  \label{restranf}
\end{align}
This invariance can easily be proven by using the following renormalization-group identity:
\begin{equation}
  h(\as(b_0^2/b^2)) = 
  h(\as(M^2)) \; \exp \left\{ 
    -\int_{b_0^2/b^2}^{M^2} \frac{\rd q^2}{q^2} 
    \;\beta(\as(q^2)) \;\frac{\rd \ln h(\as(q^2))}{\rd \ln \as(q^2)} 
  \right\} \;,
  \label{rgidenlean}
\end{equation}
which is valid for any perturbative function $h(\as)$. Notice that Eq.~\eqref{rgidenlean} establishes the evolution of the perturbative functions from the scale $q^{2}=b_0^2/b^2$ to $q^{2}=M^2$. The QCD $\beta$-function and its corresponding $n$-th order coefficient $\beta_n$ are defined as
\begin{equation}
  \frac{\rd \ln \as(\mu^2)}{\rd \ln \mu^2} = 
  \beta(\as(\mu^2)) = 
  - \sum_{n=0}^{+\infty} \beta_n \left( \frac{\as}{\pi} \right)^{n+1} \;.
  \label{rge}
\end{equation}
The explicit expression of the first three coefficients~\cite{Tarasov:1980au,Larin:1993tp}, $\beta_0$, $\beta_1$ and $\beta_2$ read
\begin{align}
  \beta_0 &= 
  \frac{1}{12} \left( 11 C_A - 2 N_f \right) \;,
  \qquad 
  \beta_1 =  
  \frac{1}{24} \left( 17 C_A^2 - 5 C_A N_f - 3 C_F N_f \right) \;,
  \nn\\
  \beta_2 &= 
  \frac{1}{64} \left( \frac{2857}{54} C_A^3
  - \frac{1415}{54} C_A^2 N_f - \frac{205}{18} C_A C_F N_f + C_F^2 N_f
  + \frac{79}{54} C_A N_f^2 + \frac{11}{9} C_F N_f^2 \right) \;,
\end{align}
where $N_f$ is the number of massless QCD flavours and the $SU(N_c)$ colour factors are $C_A=N_c$  and $C_F=(N_c^2-1)/(2N_c)$.

Throughout this paper we always use the \textit{hard resummation scheme}~\cite{Catani:2013tia} to report explicit expressions for the perturbative expansion of these individual coefficients. The \textit{hard resummation scheme} states that all the contributions proportional to $\delta(1-z)$ are associated with the hard--virtual functions $H_c^{F}$. This directly implies that $H_c^{F}$ is process dependent. 
The collinear $C_{ab}$ and $G_{ab}$ functions and the resummation coefficients $A_{c}$ and $B_{c}$ are independent of the final state system $F$. 

The truncation of Eq.~\eqref{reslean} at a given fixed order requires the explicit knowledge of resummation coefficients and hard collinear coefficient functions.
For $F=H$ at NLO, the knowledge of the coefficients $A^{(1)}_{g}$, $B^{(1)}_{g}$, $C^{(1)}_{ga}$ ($a=q,{\bar q},g$) and $H^{H;(1)}_{g}$ are sufficient to compute the inclusive total cross section and differential distributions. Assuming that the Higgs boson couples to a single heavy quark of mass $m_Q$, the first-order coefficient $H_g^{H;(1)}$ in the hard resummation scheme is~\cite{Catani:2013tia}
\begin{equation}
  H_g^{H;(1)} = C_A\pi^2/2+c_H(m_Q) \;.
  \label{H1g}
\end{equation}
The function $c_H(m_Q)$, which depends on the NLO virtual corrections of the Born subprocess, is given in Eq.~(B.2) of Ref.~\cite{Spira:1995rr}. In the limit $m_Q\to \infty$, the function $c_H$ becomes
\begin{equation}
  c_H(m_Q)\longrightarrow\frac{5C_A-3C_F}{2} \;.
\end{equation}
Therefore, the complete set of coefficients necessary to compute Higgs boson production (in the limit in which the mass of the top quark $Q=t$ is larger than any other scale involved in the process) at NLO are
\begin{align}
  A^{(1)}_{g} &= C_{A}\;, &
  B^{(1)}_{g} &= -\frac{1}{6} \left( 11 C_A - 2 N_f \right)\;, &
  H^{H;(1)}_{g} &= \frac{1}{2}(C_{A}(\pi^{2}+5)-3 C_{F})\;,
  \nn\\
  C^{(1)}_{gg}(z) &= 0 \;, &
  C^{(1)}_{ga}(z) &= \frac{1}{2}C_F\,z \qquad\left[a=q,{\bar q}\right] \;.
  \label{eq:nloconstants}
\end{align}
The coefficients $A^{(1)}_{g}$ and $B^{(1)}_{g}$ are process \emph{and} resummation scheme independent. The collinear functions $C^{(1)}_{ga}$ ($a=q,{\bar q},g$) are process independent, while $H^{H;(1)}_{g}$ depends on the final-state system $(F=H)$. Together,  they depend on the resummation scheme in such a way to ensure the resummation scheme independence of Eq.~\eqref{reslean} at NLO. In Ref.~\cite{deFlorian:2001zd} was shown that the NLO hard--virtual coefficient $H^{F;(1)}_{c}$ is explicitly related to $\rd\hat{\sigma}^{F}_{\rm LO}$ and to the IR finite part of the NLO virtual correction to the Born cross section. 

At NNLO, the coefficients $A^{(2)}_{g}$ and $B^{(2)}_{g}$ are needed~\cite{Bozzi:2005wk,Catani:2013tia,deFlorian:2001zd},
\begin{align}
  A^{(2)}_{g} &= 
  \frac{1}{2}\; C_A \left[ 
    \left( \frac{67}{18} - \frac{\pi^2}{6}\right) C_A 
    -\frac{5}{9} N_f 
  \right] \;, &
  B^{(2)}_{g} &=
  \frac{\gamma_{g}^{(1)}}{16}+\beta_0\, C_A\,\zeta_2 \;,
  \label{A2andB2}
\end{align}
where $\gamma_{g}^{(1)}$ is the coefficient of the $\delta(1-z)$ term in the NLO gluon splitting function~\cite{Curci:1980uw,Furmanski:1980cm}, which reads
\begin{equation}
  \gamma_{g}^{(1)} = 
  \left(-\frac{64}3-24\zeta_3\right)\,C_A^2
  +\frac{16}3\,C_A N_f
  +4\,C_F N_f\;.
  \label{ga1g}
\end{equation}
The coefficient $A^{(2)}_{g}$ does not depend on the resummation scheme whereas $B^{(2)}_{g}$  in Eq.~\eqref{A2andB2} is valid in the hard resummation scheme and both coefficients are process independent. 

The general structure of the hard--virtual coefficients $H^{F}_{c}$ has been established in Ref.~\cite{Catani:2013tia}. Although $H^{F}_{c}$ is in principle process dependent, Ref.~\cite{Catani:2013tia} showed it can be directly related in a universal way to the IR finite part of the all-order virtual amplitude of the corresponding partonic subprocess $c{\bar c}\to F$. The relationship between $H^{F}_{c}$ and the all-order virtual correction to the  partonic subprocess $c{\bar c}\to F$ has been made explicit up to NNLO and is based on the definition of universal subtraction operators that cancel the IR divergences of the two-loop (NNLO) virtual corrections to the Born cross section~\cite{Catani:1998bh}. These universal second-order operators contain an IR finite term of soft origin ($\delta^{(1)}_{\qt}$) that only depends on the initial-state partons~\cite{Catani:2013tia}.

In the case of Higgs boson production, the hard--virtual factor $H^{F=H;(2)}_{g}$ in the large-$m_t$ limit (in the hard resummation scheme) is given by~\cite{Catani:2011kr}
\begin{align}
  H_g^{H;(2)} &=
  C_A^2 \left(
    \frac{3187}{288}+\frac{7}{8}L_t+\frac{157}{72}\pi^2+\frac{13}{144}\pi^4-\frac{55}{18}\zeta_3
  \right)
  +C_A\, C_F \left( -\frac{145}{24}-\frac{11}{8}L_t-\frac{3}{4}\pi^2\right)
  \nn\\&\quad
  +\frac{9}{4}C_F^2 -\frac{5}{96}C_A-\frac{1}{12}C_F-C_A\, N_f\left(\frac{287}{144}+\frac{5}{36}\pi^2+\frac{4}{9}\zeta_3\right)
  +C_F\, N_f\left(-\frac{41}{24}+\frac{1}{2}L_t+\zeta_3\right) \;,
  \label{H2g}
\end{align}
where $L_t=\ln (M^2/m_t^2)$. The two-loop scattering amplitude~\cite{Harlander:2009bw} used in the computation of $H^{F=H;(2)}_{g}$ includes corrections to the large-$m_t$ approximation.

Due to the large size of the expressions for $C_{ab}^{(2)}(z)$, we refrain from explicitly quoting them here and instead refer to Eqs.~(37)--(40) of Ref.~\cite{Catani:2013tia} using the full results of Refs.~\cite{Catani:2011kr,Catani:2012qa}. These collinear coefficients $C^{(2)}_{ab}$ have been independently computed in Refs.~\cite{Gehrmann:2012ze,Echevarria:2016scs}.

At NNLO, in Eq.~\eqref{whathlean} the first order $G_{ga}^{(1)}$ helicity-flip functions are required which read~\cite{Catani:2010pd}
\begin{equation}
  G_{g \,a}^{(1)}(z) = C_a \;\frac{1-z}{z} \qquad a=q,\bar{q},g \; ,
\end{equation}
where $C_{q;\bar{q}}=C_{F}$ and $C_g=C_{A}$. The first-order functions $G_{ga}^{(1)}$ are resummation-scheme independent and do not depend on the final-state system $F$.

At \N3\LO, the numerical implementation of Eq.~\eqref{reslean} requires the following ingredients: $A^{(3)}_{g}$, $B^{(3)}_{g}$, $C^{(3)}_{ga}$, $G^{(2)}_{ga}$ ($a=q,{\bar q},g$) and $H^{H;(3)}_{g}$. The coefficient $A_g^{(3)}$~\cite{Becher:2010tm} reads
\begin{align}
  A_g^{(3)} &=  
  C_A^3 \left( 
    \frac{245}{96} 
    - \frac{67}{36}\zeta_2
    + \frac{11}{24}\zeta_3 + \frac{11}{20}\zeta_2^2\right) 
    + C_A C_F N_f \left(-\frac{55}{96} 
    + \frac{1}{2}\zeta_3
  \right)
  -C_A N_f^2 \frac{1}{108} 
  \nn\\&\quad
  + C_A^2 N_f \left(
    - \frac{209}{432} 
    + \frac{5}{18}\zeta_2 
    - \frac{7}{12} \zeta_3
  \right)  
  + \beta_{0} C_A^2 \left(
    \frac{101}{27}
    -\frac{7}{2}\zeta_{3}
  \right)
  -\beta_{0} C_A N_f \frac{14}{27} \;.
  \label{acoeff}
\end{align}
The explicit expression of the $B_c^{(3)}$ ($a=q,g$) coefficients in the hard scheme can be computed from Refs.~\cite{Li:2016ctv,Vladimirov:2016dll}. In the particular case of the gluon channel then in the hard resummation scheme, we obtain
\begin{align}
  B_g^{(3)} &= 
  - \frac{2133}{64} 
  + \frac{3029}{576} N_{f} 
  - \frac{349}{1728} N_{f}^{2}  
  + \frac{109}{6} \pi^{2} 
  - \frac{283}{144} \pi^{2}  N_{f} 
  + \frac{5}{108} \pi^{2}  N_{f}^{2}
  -\frac{253}{160} \pi^{4}
  \nn\\&\quad
  +\frac{23}{240} \pi^{4} N_{f}
  -\frac{843}{8} \zeta_{3} 
  + 2 \zeta_{3} N_{f} 
  + \frac{1}{6} \zeta_{3} N_{f}^{2} 
  + \frac{9}{4} \pi^{2}  \zeta_{3} 
  + \frac{135}{2}  \zeta_{5} \;.
\end{align}

\section{Convolutions at \texorpdfstring{N${}^\text{3}$LO}{N3LO}}
\label{app:Convos}

The numerical implementation of Eq.~\eqref{reslean} requires the computation of several convolutions between splitting functions, collinear and helicity-flip functions. In principle, taking the $N$-moments of the functions involved in the calculation, one can avoid the use of convolutions, since in $N$-space they correspond to simple products. However, the numerical implementation of Eq.~\eqref{reslean} in the Monte Carlo code \texttt{HN3LO} was carried out in the $z$-space (e.g.\ as in the codes \texttt{HNNLO}~\cite{Catani:2007vq}, \texttt{DYNNLO}~\cite{Catani:2009sm}, \texttt{2$\gamma$NNLO}~\cite{Catani:2011qz}, etc.), and therefore the new third order convolutions have to be calculated as well. 

The convolutions in Eqs.~\eqref{H2}, \eqref{H3}, \eqref{CN3eq} and \eqref{CNeq} between two functions ($f(z)$ and $g(z)$) of the the variable $z$ are defined through the following integral
\begin{align}
  \left( f \otimes g \right)(z) &\equiv \int^{1}_{z} \; \frac{\rd y}{y} \; f\left(\frac{z}{y}\right) \; g(y) \;.
\end{align}
In the case of processes initiated by gluon fusion, the complete list of third order convolutions to be calculated can be found in Table~\ref{Table:convosN3LO}. All the remaining convolutions in Eq.~\eqref{reslean} at \N3\LO already  contributed to the previous orders and they are regarded as known.

\begin{table}
\centering
\renewcommand{\arraystretch}{1.5}
\begin{tabular}{ |c|c||c|c| }
\hline
\multirow{1}{*}{(i)} 

& $\gamma^{(1)}_{ga}\otimes\gamma^{(1)}_{a b}\otimes\gamma^{(1)}_{bg}$

& \multirow{1}{*}{(ii)} 

& $\gamma^{(1)}_{ga}\otimes\gamma^{(1)}_{ab}\otimes\gamma^{(1)}_{bq}$  \\

\multirow{1}{*}{(iii)} 

& $\gamma^{(1)}_{ga}\otimes\gamma^{(2)}_{ag}$

& \multirow{1}{*}{(iv)} 

& $\gamma^{(1)}_{ga}\otimes\gamma^{(2)}_{aq}$  \\

\multirow{1}{*}{(v)} 

& $\gamma^{(2)}_{ga}\otimes\gamma^{(1)}_{ag}$

& \multirow{1}{*}{(vi)} 

& $\gamma^{(2)}_{ga}\otimes\gamma^{(1)}_{aq}$  \\

\multirow{1}{*}{(vii)} 

& $C^{(1)}_{ga}\otimes\gamma^{(2)}_{ag}$

& \multirow{1}{*}{(viii)} 

& $C^{(1)}_{ga}\otimes\gamma^{(2)}_{aq}$  \\

\multirow{1}{*}{(ix)} 

& $C^{(2)}_{ga}\otimes\gamma^{(1)}_{ag}$

& \multirow{1}{*}{(x)} 

& $C^{(2)}_{ga}\otimes\gamma^{(1)}_{aq}$  \\

\multirow{1}{*}{(xi)} 

& $G^{(1)}_{ga}\otimes\gamma^{(1)}_{ag}$

& \multirow{1}{*}{(xii)} 

& $G^{(1)}_{ga}\otimes\gamma^{(1)}_{aq}$  \\

\hline
\end{tabular}
\caption{\label{Table:convosN3LO}
{Convolutions appearing at the \N3\LO-only between the collinear $C^{(n)}_{ab}$, the helicity-flip $G^{(n)}_{ab}$ and the splitting functions $\gamma^{(n)}_{ab}$ ($n=1,2)$. The repeated subindices $a$ and $b$ imply a sum over the parton flavors $q,\bar{q},g$.  The first and last subindices denote the partonic channel in which they are contributing, i.e.\ the convolutions in the first column are used in the $gg$ partonic channel whereas the second (and last) column is for the $qg$ and $gq$ partonic channels.
}}
\renewcommand{\arraystretch}{1}
\end{table}

The symbol $\gamma_{ab}^{(n)}$ in Table~\ref{Table:convosN3LO} denotes the usual splitting functions of $n$-th order and they contribute to Eq.~\eqref{reslean}) since the PDFs have to be evolved from the scale $b_0^2/b^2$ to the factorization scale $\muF$.
The first three rows in Eq.~\eqref{Table:convosN3LO} were calculated in Ref.~\cite{Hoeschele:2012xc} and cross-checked with a dedicated computation for the results presented in this paper. The public \texttt{Mathematica} package \texttt{MT}~\cite{Hoeschele:2013gga} is used to calculate the necessary convolutions (i)--(vi) in Ref.~\cite{Hoeschele:2012xc}, which can be further expressed in terms of \textit{harmonic polylogarithms} (HPLs)~\cite{Remiddi:1999ew} using the \texttt{Mathematica} package \texttt{HPL}~\cite{Maitre:2005uu}. 
The remaining convolutions in Eqs. (vii)--(xii) of Table~\ref{Table:convosN3LO}  were computed  for this work. The \texttt{MT}~\cite{Hoeschele:2013gga} package is not able to solve all the convolutions of weight 3 and 4 that are needed in (vii)--(xii). For instance, the \texttt{MT} package cannot handle convolutions in which their result has to be expressed in terms of multiple polylogarithms (or \emph{Goncharov polylogarithms} GPLs)~\cite{Goncharov:1998kja,Gehrmann:2000zt,Goncharov:2001iea} as it is the case when the collinear functions $C^{(2)}_{gj}$ are involved. For those, we have computed the convolutions (vii)--(xii) with a newly developed code \texttt{Convo}, which is able to provide results in terms of GPLs and also can handle terms that are individually divergent, but finite after addition.

The multiple polylogarithms can be defined recursively, for $n\geq 0$, via the iterated integral~\cite{Goncharov:1998kja,Gehrmann:2000zt,Goncharov:2001iea}
\begin{align}
   G(a_1,\ldots,a_n;z) &= \,\int_0^z\,\frac{\rd t}{t-a_1}\,G(a_2,\ldots,a_n;t) \;,
  \label{MultPolyLogdef}
\end{align}
with $G(z) = G(;z)=1$ (an exception being when $z=0$ in which case we put $G(0)=0$) and with $a_i\in \mathbb{C}$ are chosen constants and $z$ is a complex variable. For the convolutions in Table~\ref{Table:convosN3LO} the variable $z$ and the weights $a_1,\ldots,a_n$ are all real constants.

From the convolutions in Table~\ref{Table:convosN3LO} we quote some  examples which appear as building blocks in the computation of Eqs.~(vii)--(xii),
\begin{align}
  \bigg\{\mathrm{D}_{0}[1-y];\frac{1}{y};1;y;y^{2}\bigg\} \otimes \left( \frac{f(y)}{1+y} \right) \;,
\end{align}
with 
\begin{align}
  f(y) =
  \bigg\{{\rm Li}_{3}\left(\frac{1}{1+y}\right); {\rm Li}_{3}(\pm y); 
  & {\rm Li}_{2}(\pm y); {\rm Li}_{2}(1- y); {\rm Li}_{2}(\pm y) \ln(y);
  \nn\\&\quad
  \ln^{2}(1+y)\ln(y);\ln(1+y)\ln^{2}(y)\bigg\} \;,
\end{align}
where the \textit{plus} distribution ${\rm D}_{0}[1-z]$ is defined as usual 
\begin{align}
  \int^{1}_{0}\rd z \; f(z) \; {\rm D}_{0}[1-z] =\int^{1}_{0}\rd z \; \frac{f(z)}{(1-z)_{+}} = \int^{1}_{0}\frac{\rd z}{1-z}\left( f(z)-f(1)\right) \;.
\end{align}
After performing all the convolutions listed in Table~\ref{Table:convosN3LO}, their final expressions (each one of the convolutions) are finite in the domain $z\in (0,1)$. Even more, convolutions evaluated in the domain $z\in (0,1)$ produce results in $\mathbb{R}$. It is possible to write the expressions in Table~\ref{Table:convosN3LO} (after simplifying) in terms of twelve GPLs that are not reducible to polylogarithmic functions of type ${\rm Li}_n(z)$, 
 and cannot be combined (e.g.\ through the \textit{shuffle} algebra) with other GPLs in order to produce simpler results. 
The list of the irreducible GPLs is presented in Table~\ref{Table:GPLsused}.
\begin{table}
\begin{center}
\renewcommand{\arraystretch}{1.5}
\begin{tabular}{ |c|c||c|c||c|c| }
\hline
\multirow{1}{*}{(a)} 

& $G(\frac{z}{1+z},0,0,1;\frac{1}{2})$

& \multirow{1}{*}{(b)} 

& $G(1,0,0,-z;z)$ 

& \multirow{1}{*}{(c)} 

& $G(0,1,0,-1;z)$  \\

\multirow{1}{*}{(d)} 

& $G(0,1,0,z;1)$



& \multirow{1}{*}{(e)} 

& $G(0,1,z,0;1)$  

& \multirow{1}{*}{(f)} 

& $G(0,z,1,0;1)$\\

\multirow{1}{*}{(g)} 

& $G(-z,0,z,0;1)$ 

& \multirow{1}{*}{(h)} 

& $G(0,1,0,-z;z)$  

&\multirow{1}{*}{(i)} 

& $G(0,1,-z,-z;z)$\\

 \multirow{1}{*}{(j)} 

& $G(-z,1,0,0;1)$ 

& \multirow{1}{*}{(k)} 

& $G(-z,1,0,0;z)$  

& \multirow{1}{*}{(l)} 

& $G(-z,0,0,z;1)$

\\  

\hline
\end{tabular}
\caption{\label{Table:GPLsused}
{Basis for the GPLs used in the numerical implementation of the convolutions listed in Table~\ref{Table:convosN3LO}. 
}}
\renewcommand{\arraystretch}{1}
\end{center}
\end{table}
All remaining GPLs appearing in the convolutions of Table~\ref{Table:convosN3LO} can be related to the set given in Table~\ref{Table:GPLsused} using the results of Refs.~\cite{Frellesvig:2016ske,Maitre:2005uu,Duhr:2011zq} and performing the customary \textit{shuffle} algebra. The numerical implementation of the GPLs in Table~\ref{Table:GPLsused} was made using the package \texttt{GiNaC}~\cite{Bauer:2000cp,Vollinga:2004sn}. The basis of GPLs in Table~\ref{Table:GPLsused} is not unique, but 
sufficient for  numerical evaluation.

An example of a third order convolution is the following integral
\begin{align}
  &\left(\frac{{\rm Li}_{3}( y) }{1+y}\otimes {\rm D}_{0}[1-y]\right)(z) 
  = 
  \int^{1}_{z}~\frac{\rd y}{y+z}~{\rm Li}_{3}\left(\frac{z}{y}\right)~\frac{1}{(1-y)_{+}}  
  \nn\\&\qquad
  = \frac{1}{1+z}\bigg(-  \zeta_{3} G(0;z)+\frac{\ri \pi ^3 }{6 }G(0;z)
  +\frac{\pi ^2}{3}G(-z;1) G(0;z)- \ri \pi  G(-z,0;1)G(0;z)
  \nn\\&\qquad\qquad
  -G(-z,0,0;1) G(0;z)
  +\frac{\ri\pi  \zeta_{3}}{4}+\frac{\pi ^2 }{3 }G(0,1;z)
   +\ri \pi  G(-z;1) G(0,0;z)-\frac{\pi ^2 }{6 }G(0,0;z)
  \nn\\&\qquad\qquad
  -G(-z;1) G(0,0,0;z)+\ri \pi G(0,0,1;z)
  +G(0,0;z)G(-z,0;1)-\frac{\pi ^2 }{3}G(-z,0;1)
  \nn\\&\qquad\qquad
  +\ri \pi G(-z,0,0;1)-G(0,0,0,1;z)
  -G(0,0,1,z;1)-G(0,0,z,1;1)-G(0,1,0,z;1)
  \nn\\&\qquad\qquad
  -G(1,0,0,z;z)+G(-z,0,0,0;1)
  -G(-z,0,0,z;1)+G(-z,0,0,z;z) +\frac{19 \pi ^4}{720 }\bigg) \;.
  \label{Convoexamp}
\end{align}

\end{appendix}

\newpage


\begin{thebibliography}{99}
\bibitem{Catani:2007vq}
  S.~Catani and M.~Grazzini,
  Phys.\ Rev.\ Lett.\  {\bf 98} (2007) 222002
  [hep-ph/0703012].
%
\bibitem{Bozzi:2005wk}
  G.~Bozzi, S.~Catani, D.~de Florian and M.~Grazzini,
  Nucl.\ Phys.\ B {\bf 737} (2006) 73
  [hep-ph/0508068].
%
\bibitem{Bonciani:2015sha}
  R.~Bonciani, S.~Catani, M.~Grazzini, H.~Sargsyan and A.~Torre,
  Eur.\ Phys.\ J.\ C {\bf 75} (2015)   581
  [arXiv:1508.03585].




\bibitem{Boughezal:2015eha}
  R.~Boughezal, X.~Liu and F.~Petriello,
  Phys.\ Rev.\ D {\bf 91} (2015) 094035
  [arXiv:1504.02540].

\bibitem{Gaunt:2015pea}
  J.~Gaunt, M.~Stahlhofen, F.~J.~Tackmann and J.~R.~Walsh,
  JHEP {\bf 1509} (2015) 058
  [arXiv:1505.04794].

\bibitem{Czakon:2011ve}
  M.~Czakon,
  Nucl.\ Phys.\ B {\bf 849} (2011) 250
  [arXiv:1101.0642].

\bibitem{Boughezal:2011jf}
  R.~Boughezal, K.~Melnikov and F.~Petriello,
  Phys.\ Rev.\ D {\bf 85} (2012) 034025
  [arXiv:1111.7041].


\bibitem{Cacciari:2015jma}
  M.~Cacciari, F.~A.~Dreyer, A.~Karlberg, G.~P.~Salam and G.~Zanderighi,
  Phys.\ Rev.\ Lett.\  {\bf 115} (2015) 082002
  Erratum: [Phys.\ Rev.\ Lett.\  {\bf 120} (2018)  139901]
  [arXiv:1506.02660].



\bibitem{Antenna:method}
  A.~Gehrmann-De Ridder, T.~Gehrmann and E.~W.~N.~Glover,
  JHEP {\bf 0509} (2005) 056
  [hep-ph/0505111];
  A.~Daleo, T.~Gehrmann and D.~Maitre,
  JHEP {\bf 0704} (2007) 016
  [hep-ph/0612257];
  J.~Currie, E.~W.~N.~Glover and S.~Wells,
  JHEP {\bf 1304} (2013) 066
  [arXiv:1301.4693].
  






\bibitem{Chetyrkin:1994js}
  K.~G.~Chetyrkin, J.~H.~Kuhn and A.~Kwiatkowski,
  Phys.\ Rept.\  {\bf 277} (1996) 189
   [hep-ph/9503396].


\bibitem{Vermaseren:2005qc}
  J.~A.~M.~Vermaseren, A.~Vogt and S.~Moch,
  Nucl.\ Phys.\ B {\bf 724} (2005) 3
  [hep-ph/0504242].

\bibitem{Anastasiou:2015ema}
  C.~Anastasiou, C.~Duhr, F.~Dulat, F.~Herzog and B.~Mistlberger,
  Phys.\ Rev.\ Lett.\  {\bf 114} (2015) 212001
   [arXiv:1503.06056].


\bibitem{Mistlberger:2018etf}
  B.~Mistlberger,
  JHEP {\bf 1805} (2018) 028
   [arXiv:1802.00833].


\bibitem{Dreyer:2016oyx}
  F.~A.~Dreyer and A.~Karlberg,
  Phys.\ Rev.\ Lett.\  {\bf 117} (2016) 072001
  [arXiv:1606.00840].

\bibitem{Dulat:2017prg}
  F.~Dulat, B.~Mistlberger and A.~Pelloni,
  JHEP {\bf 1801} (2018) 145
   [arXiv:1710.03016].
  
\bibitem{Dulat:2018bfe}
  F.~Dulat, B.~Mistlberger and A.~Pelloni,
  arXiv:1810.09462 [hep-ph].
  
\bibitem{Currie:2018fgr}
  J.~Currie, T.~Gehrmann, E.~W.~N.~Glover, A.~Huss, J.~Niehues and A.~Vogt,
  JHEP {\bf 1805} (2018) 209
   [arXiv:1803.09973].

  


  
  \bibitem{qTRes:program}
Y.~L.~Dokshitzer, D.~Diakonov and S.~I.~Troian,
Phys.\ Lett.\  B {\bf 79} (1978) 269,
Phys.\ Rep.\  {\bf 58} (1980) 269;
G.~Parisi and R.~Petronzio,
Nucl.\ Phys.\ B {\bf 154} (1979) 427.
G.~Curci, M.~Greco and Y.~Srivastava,
Nucl.\ Phys.\ B {\bf 159} (1979) 451;
J.~C.~Collins and D.~E.~Soper,
Nucl.\ Phys.\ B {\bf 193} (1981) 381
[Erratum-ibid.\ B {\bf 213} (1983) 545],
Nucl.\ Phys.\ B {\bf 197} (1982) 446;
J.~Kodaira and L.~Trentadue,
Phys.\ Lett.\ B {\bf 112} (1982) 66,
report SLAC-PUB-2934 (1982),
Phys.\ Lett.\ B {\bf 123} (1983) 335;
J.~C.~Collins, D.~E.~Soper and G.~Sterman,
Nucl.\ Phys.\ B {\bf 250} (1985) 199;
S.~Catani, E.~D'Emilio and L.~Trentadue,
Phys.\ Lett.\ B {\bf 211} (1988) 335;
  D.~de Florian and M.~Grazzini,
  Phys.\ Rev.\ Lett.\  {\bf 85} (2000) 4678
[hep-ph/0008152];
  S.~Catani, D.~de Florian and M.~Grazzini,
  Nucl.\ Phys.\  B {\bf 596} (2001) 299
[hep-ph/0008184].

\bibitem{Bizon:2017rah}
  W.~Bizon, P.~F.~Monni, E.~Re, L.~Rottoli and P.~Torrielli,
  JHEP {\bf 1802} (2018) 108
    [arXiv:1705.09127].

  
\bibitem{Catani:2010pd}
  S.~Catani and M.~Grazzini,
  Nucl.\ Phys.\ B {\bf 845} (2011) 297
[arXiv:1011.3918].

\bibitem{Catani:2013tia}
  S.~Catani, L.~Cieri, D.~de Florian, G.~Ferrera and M.~Grazzini,
  Nucl.\ Phys.\ B {\bf 881} (2014) 414
   [arXiv:1311.1654].
  
\bibitem{Catani:2000vq}
  S.~Catani, D.~de Florian and M.~Grazzini,
  Nucl.\ Phys.\  B {\bf 596} (2001) 299
[hep-ph/0008184].

  \bibitem{Tarasov:1980au}
  O.~V.~Tarasov, A.~A.~Vladimirov and A.~Y.~Zharkov,
  Phys.\ Lett.\  {\bf 93B} (1980) 429.
  
\bibitem{Larin:1993tp}
  S.~A.~Larin and J.~A.~M.~Vermaseren,
  Phys.\ Lett.\ B {\bf 303} (1993) 334
  [hep-ph/9302208].



\bibitem{deFlorian:2001zd}
  D.~de Florian and M.~Grazzini,
  Nucl.\ Phys.\  B {\bf 616} (2001) 247
[hep-ph/0108273].
 
\bibitem{Catani:2011kr}
  S.~Catani and M.~Grazzini,
  Eur.\ Phys.\ J.\ C {\bf 72} (2012) 2013
[Erratum-ibid.\ C {\bf 72} (2012) 2132]
[arXiv:1106.4652].


\bibitem{Catani:2012qa}
  S.~Catani, L.~Cieri, D.~de Florian, G.~Ferrera and M.~Grazzini,
  Eur.\ Phys.\ J.\ C {\bf 72} (2012) 2195
  [arXiv:1209.0158].

\bibitem{Gehrmann:2012ze}
   T.~Gehrmann, T.~L\"ubbert and L.~L.~Yang,
   Phys.\ Rev.\ Lett.\  {\bf 109} (2012) 242003
   [arXiv:1209.0682];
   JHEP {\bf 1406} (2014) 155
   [arXiv:1403.6451].

\bibitem{Echevarria:2016scs}
  M.~G.~Echevarria, I.~Scimemi and A.~Vladimirov,
  JHEP {\bf 1609} (2016) 004
  [arXiv:1604.07869].

  

\bibitem{Curci:1980uw}
  G.~Curci, W.~Furmanski and R.~Petronzio,
  Nucl.\ Phys.\ B {\bf 175} (1980) 27.

\bibitem{Furmanski:1980cm}
  W.~Furmanski and R.~Petronzio,
  Phys.\ Lett.\ B {\bf 97} (1980) 437.


\bibitem{Harlander:2009bw}
  R.~V.~Harlander and K.~J.~Ozeren,
  Phys.\ Lett.\ B {\bf 679} (2009) 467
  [arXiv:0907.2997];



\bibitem{Catani:1998bh}
  S.~Catani,
  Phys.\ Lett.\ B {\bf 427} (1998) 161
  [hep-ph/9802439].



  
\bibitem{Catani:1988vd}
  S.~Catani, E.~D'Emilio and L.~Trentadue,
  Phys.\ Lett.\ B {\bf 211} (1988) 335.

\bibitem{Kauffman:1991cx}
  R.~P.~Kauffman,
  Phys.\ Rev.\ D {\bf 45} (1992) 1512.

\bibitem{deFlorian:2000pr}
  D.~de Florian and M.~Grazzini,
  Phys.\ Rev.\ Lett.\  {\bf 85} (2000) 4678
  [hep-ph/0008152];
    D.~de Florian, G.~Ferrera, M.~Grazzini and D.~Tommasini,
  JHEP {\bf 1111} (2011) 064
  [arXiv:1109.2109].

  
\bibitem{Becher:2012yn}
  T.~Becher, M.~Neubert and D.~Wilhelm,
  JHEP {\bf 1305} (2013) 110
  [arXiv:1212.2621].

\bibitem{Neill:2015roa}
  D.~Neill, I.~Z.~Rothstein and V.~Vaidya,
  JHEP {\bf 1512} (2015) 097
  [arXiv:1503.00005].

  
\bibitem{Chen:2018pzu}
  X.~Chen {\it et al.},
  Phys.\ Lett.\ B {\bf 788} (2019) 425
  [arXiv:1805.00736].

\bibitem{Bizon:2018foh}
  W.~Bizon {\it et al.},
  arXiv:1805.05916.

  

\bibitem{Spira:1995rr}
  M.~Spira, A.~Djouadi, D.~Graudenz and P.~M.~Zerwas,
  Nucl.\ Phys.\ B {\bf 453} (1995) 17
  [hep-ph/9504378].

\bibitem{Li:2016ctv}
  Y.~Li and H.~X.~Zhu,
  Phys.\ Rev.\ Lett.\  {\bf 118} (2017)  022004
   [arXiv:1604.01404].
  
\bibitem{Vladimirov:2016dll}
  A.~A.~Vladimirov,
  Phys.\ Rev.\ Lett.\  {\bf 118} (2017),  062001
   [arXiv:1610.05791].
  
\bibitem{Catani:2014uta}
  S.~Catani, L.~Cieri, D.~de Florian, G.~Ferrera and M.~Grazzini,
  Nucl.\ Phys.\ B {\bf 888} (2014) 75
  [arXiv:1405.4827].
  

\bibitem{Becher:2010tm}
  T.~Becher, M.~Neubert,
  Eur.\ Phys.\ J.\  {\bf C71 } (2011)  1665
[arXiv:1007.4005].


  \bibitem{Heft}
  F.~Wilczek,
  Phys.\ Rev.\ Lett.\  {\bf 39} (1977) 1304;\\
 M.~A.~Shifman, A.~I.~Vainshtein and V.~I.~Zakharov,
  Phys.\ Lett.\ B {\bf 78} (1978) 443;\\
  T.~Inami, T.~Kubota and Y.~Okada,
  Z.\ Phys.\ C {\bf 18} (1983) 69.

  
\bibitem{nnpdf}
   R.~D.~Ball {\it et al.} [NNPDF Collaboration],
  JHEP {\bf 1504} (2015) 040
  [arXiv:1410.8849].

\bibitem{Buckley:2014ana}
  A.~Buckley, J.~Ferrando, S.~Lloyd, K.~Nordstr\"om, B.~Page, M.~R\"ufenacht, M.~Sch\"onherr and G.~Watt,
  Eur.\ Phys.\ J.\ C {\bf 75} (2015) 132
  [arXiv:1412.7420].

\bibitem{Chen:2016zka}
  X.~Chen, J.~Cruz-Martinez, T.~Gehrmann, E.~W.~N.~Glover and M.~Jaquier,
  JHEP {\bf 1610} (2016) 066
  [arXiv:1607.08817].

\bibitem{Ridder:2015dxa}
  A.~Gehrmann-De Ridder, T.~Gehrmann, E.~W.~N.~Glover, A.~Huss and T.~A.~Morgan,
  Phys.\ Rev.\ Lett.\  {\bf 117} (2016)   022001
  [arXiv:1507.02850].

\bibitem{Gehrmann-DeRidder:2016jns}
  A.~Gehrmann-De Ridder, T.~Gehrmann, E.~W.~N.~Glover, A.~Huss and T.~A.~Morgan,
  JHEP {\bf 1611} (2016) 094
  [arXiv:1610.01843].
  
\bibitem{Gehrmann-DeRidder:2017mvr}
  A.~Gehrmann-De Ridder, T.~Gehrmann, E.~W.~N.~Glover, A.~Huss and D.~M.~Walker,
  Phys.\ Rev.\ Lett.\  {\bf 120} (2018)  122001
  [arXiv:1712.07543].

\bibitem{Gehrmann:2011aa}
  T.~Gehrmann, M.~Jaquier, E.~W.~N.~Glover and A.~Koukoutsakis,
  JHEP {\bf 1202} (2012) 056
   [arXiv:1112.3554].

\bibitem{Dixon:2009uk}
  L.~J.~Dixon and Y.~Sofianatos,
  JHEP {\bf 0908} (2009) 058
   [arXiv:0906.0008].


\bibitem{Badger:2009hw}
  S.~Badger, E.~W.~N.~Glover, P.~Mastrolia and C.~Williams,
  JHEP {\bf 1001} (2010) 036
   [arXiv:0909.4475].


\bibitem{Badger:2009vh}
  S.~Badger, J.~M.~Campbell, R.~K.~Ellis and C.~Williams,
  JHEP {\bf 0912} (2009) 035
  [arXiv:0910.4481].



\bibitem{DelDuca:2004wt}
  V.~Del Duca, A.~Frizzo and F.~Maltoni,
  JHEP {\bf 0405} (2004) 064
  [hep-ph/0404013].

\bibitem{Dixon:2004za}
  L.~J.~Dixon, E.~W.~N.~Glover and V.~V.~Khoze,
  JHEP {\bf 0412} (2004) 015
  [hep-th/0411092].


\bibitem{Badger:2004ty}
  S.~D.~Badger, E.~W.~N.~Glover and V.~V.~Khoze,
  JHEP {\bf 0503} (2005) 023
  [hep-th/0412275].

\bibitem{leaninprep}

L.~Cieri, In preparation. 

\bibitem{leaninprepHN3LO}

L.~Cieri, In preparation. 

\bibitem{Dulat:2018rbf}
  F.~Dulat, A.~Lazopoulos and B.~Mistlberger,
  Comput.\ Phys.\ Commun.\  {\bf 233} (2018) 243
  [arXiv:1802.00827].
  

\bibitem{Anastasiou:2016cez}
  C.~Anastasiou, C.~Duhr, F.~Dulat, E.~Furlan, T.~Gehrmann, F.~Herzog, A.~Lazopoulos and B.~Mistlberger,
  JHEP {\bf 1605} (2016) 058
   [arXiv:1602.00695].

  
  
  
\bibitem{Catani:2009sm}
  S.~Catani, L.~Cieri, G.~Ferrera, D.~de Florian and M.~Grazzini,
  Phys.\ Rev.\ Lett.\  {\bf 103} (2009) 082001
  [arXiv:0903.2120].

\bibitem{Catani:2011qz}
  S.~Catani, L.~Cieri, D.~de Florian, G.~Ferrera and M.~Grazzini,
  Phys.\ Rev.\ Lett.\  {\bf 108} (2012) 072001
   Erratum: [Phys.\ Rev.\ Lett.\  {\bf 117} (2016),  089901]
  [arXiv:1110.2375].





\bibitem{Hoeschele:2013gga}
  M.~H\"oschele, J.~Hoff, A.~Pak, M.~Steinhauser and T.~Ueda,
  Comput.\ Phys.\ Commun.\  {\bf 185} (2014) 528
  [arXiv:1307.6925].
  
  \bibitem{Hoeschele:2012xc}
  M.~H\"oschele, J.~Hoff, A.~Pak, M.~Steinhauser and T.~Ueda,
  Phys.\ Lett.\ B {\bf 721} (2013) 244
  [arXiv:1211.6559].
  
\bibitem{Gehrmann:2000zt}
  T.~Gehrmann and E.~Remiddi,
  Nucl.\ Phys.\ B {\bf 601} (2001) 248
   [hep-ph/0008287].

  
  \bibitem{Maitre:2005uu}
  D.~Maitre,
  Comput.\ Phys.\ Commun.\  {\bf 174} (2006) 222
   [hep-ph/0507152].
  
  \bibitem{Remiddi:1999ew}
  E.~Remiddi and J.~A.~M.~Vermaseren,
  Int.\ J.\ Mod.\ Phys.\ A {\bf 15} (2000) 725
   [hep-ph/9905237].
  
  \bibitem{Goncharov:1998kja}
  A.~B.~Goncharov,
  Math.\ Res.\ Lett.\  {\bf 5} (1998) 497
   [arXiv:1105.2076 [math.AG]].
  
  \bibitem{Goncharov:2001iea}
  A.~B.~Goncharov,
  math/0103059 [math.AG].
  
  \bibitem{Duhr:2011zq}
  C.~Duhr, H.~Gangl and J.~R.~Rhodes,
  JHEP {\bf 1210} (2012) 075
   [arXiv:1110.0458].
  
  \bibitem{Frellesvig:2016ske}
  H.~Frellesvig, D.~Tommasini and C.~Wever,
  JHEP {\bf 1603} (2016) 189
  [arXiv:1601.02649].
  
  \bibitem{Bauer:2000cp}
  C.~W.~Bauer, A.~Frink and R.~Kreckel,
  J.\ Symb.\ Comput.\  {\bf 33} (2000) 1
  [cs/0004015].

\bibitem{Vollinga:2004sn}
  J.~Vollinga and S.~Weinzierl,
  Comput.\ Phys.\ Commun.\  {\bf 167} (2005) 177
  [hep-ph/0410259].


\end{thebibliography}
\end{document}